\documentclass[12pt]{iopart}

\usepackage[T1]{fontenc}
\usepackage{iopams}
\expandafter\let\csname equation*\endcsname\relax
\expandafter\let\csname endequation*\endcsname\relax
\usepackage{amsmath}
\usepackage{amssymb}
\usepackage{graphicx}
\newcommand{\Ref}[1]{(\ref{#1})}
\newcommand{\dd}{\ensuremath{\mathrm{d}}}
\newcommand{\df}[2]{\ensuremath{\frac{\dd #1}{\dd #2}}}
\newcommand{\f}[2]{\ensuremath{\frac{#1}{#2}}}
\newcommand{\deq}{\ensuremath{\overset{d}{=}}}
\newcommand{\defeq}{\mathrel{\mathop:}=}

\let\originalleft\left
\let\originalright\right
\renewcommand{\left}{\mathopen{}\mathclose\bgroup\originalleft}
\renewcommand{\right}{\aftergroup\egroup\originalright}

\usepackage{xcolor}
\newcommand{\XXX}[1]{ #1}

\begin{document}

\title[Random coefficient autoregressive processes and non-Gaussian Brownian
motion]{Random coefficient autoregressive processes describe Brownian yet
non-Gaussian diffusion in heterogeneous systems}

\author{Jakub {\'S}l\k{e}zak$^{\dagger\ddagger}$, Krzysztof Burnecki$^\ddagger$
\& Ralf Metzler$^\sharp$}
\address{$^\dagger$Department of Physics, Bar Ilan University\\
$^\ddag$Faculty of Pure and Applied Mathematics, Wroc{\l}aw University of
Science and Technology\\
$^\sharp$Institute of Physics and Astronomy, Potsdam University}
\ead{rmetzler@uni-potsdam.de. Corresponding author: Ralf Metzler}

\begin{abstract}
Many studies on biological and soft matter systems report the joint presence
of a linear mean-squared displacement and a non-Gaussian probability
density exhibiting, for instance, exponential or stretched-Gaussian
tails. This phenomenon is ascribed to the heterogeneity of the medium and is
captured by random parameter models such as "superstatistics" or "diffusing
diffusivity". Independently, scientists working in the area of time series
analysis and statistics have studied  a class of discrete-time processes
with similar properties, namely, random coefficient autoregressive models. In
this work we try to reconcile these two approaches and thus provide a bridge
between physical stochastic processes and autoregressive models. We start from
the basic Langevin equation of motion with time-varying damping or diffusion
coefficients and establish the link to random coefficient autoregressive
processes. By exploring that link we gain access to efficient statistical
methods which can help to identify data exhibiting Brownian yet non-Gaussian
diffusion.
\end{abstract}

\section{Introduction}

Brownian motion, one of the most fundamental processes in non-equilibrium
statistical physics, describes the motion of a passive colloidal particle
in a thermal fluid environment. It was observed even in ancient times, for instance, by Roman philosopher Lucretius \cite{lucretius}. The modern scientific interest was started  by botanist Robert Brown \cite{brown};
it was later studied theoretically by Einstein, Sutherland, Smoluchowski, and
Langevin between 1905 and 1908 \cite{einstein,sutherland,smoluchowski,langevin}.
Two fundamental properties are typically associated with Brownian motion, namely,
the linear growth
\begin{equation}
\label{msd}
\delta_X^2(t)\defeq \mathbb{E}[ X(t)^2]=2Dt
\end{equation}
in time of the mean-squared displacement (MSD) with the diffusion coefficient $D$,
and the Gaussian probability density function (PDF)
\begin{equation}
p_X(x;t)=\f{1}{\sqrt{4\pi Dt}}\exp\left(-\f{x^2}{4Dt}\right).
\end{equation}
Alternatively to the MSD (\ref{msd}) Brownian motion is characterised by a $1/f^2$
frequency dependence of the associated ensemble and single trajectory power
spectra \cite{power,krapf}.

Deviations from the linear time dependence (\ref{msd}) of the MSD are observed
routinely in a large variety of systems, in particular, in the power-law form
\begin{equation}
\label{msd1}
\delta_X^2(t)\sim D_{\alpha}t^{\alpha}
\end{equation}
with the anomalous diffusion coefficient $D_{\alpha}$ \cite{bouchaud,report,pccp}.
In biological cells or other complex liquids both subdiffusion with $0<\alpha<1$
\cite{golding, garini_poland,lene,weigel,lene1,matthias} and superdiffusion with
$1<\alpha<2$ \cite{elbaum,christine,robert} are measured, see the recent reviews
\cite{hoefling, lene_rev}. Anomalous diffusion are effected when the increments
of the stochastic process are no longer independent, when the variance of the
step length or the mean-step time diverges, as well as due to the tortuosity of
the embedding space. The associated PDF of these processes may have both Gaussian
and non-Gaussian shapes \cite{bouchaud,report,pccp}.

Of late, a new class of diffusive dynamics has come into focus, following numerous
reports in soft matter, biological and other complex systems: in these systems the
MSD is normal of the form (\ref{msd}) with invariable coefficient $D$, however, the
PDF is non-Gaussian and often found to be of the distinct exponential shape
("Laplace distribution")
\begin{equation}
\label{eq:LaplPDF}
p_X(x;t)=\f{1}{\sqrt{2D t}}\exp\left(-\sqrt{\f{2}{Dt}}|x|\right),
\end{equation}
see \cite{wang2,wangNG,bhattacharya,hapca,beta} as well as the extensive list of
references in \cite{diffDiffChechkin}. These "Brownian yet non-Gaussian" processes
along with a more general class of non-Gaussian PDFs, discussed in more detail
below, are in the focus of this study. 

Our goal here is, however, different from
the previous modelling approaches. Namely, here we try to establish a direct
connection of the physical models for Brownian yet non-Gaussian diffusion and
a class of processes ubiquitously used in time series analysis, the so-called
autoregressive models with random coefficients \cite{tsay00,BJ,BD}.

\XXX{Thus, we introduce the time series methods to the modeling of non-Gaussian diffusion, which enables us to provide information about the distribution of the process, showing the influence of some effective properties of the heterogeneous medium. In particular it allows to distinguish between locally homogeneous (''superstastics'' type) and rapidly varying (''diffusing diffusivity'' type)  environments (see, e.g. figure \ref{fig:intPDF}). It also makes it possible to show how the heterogeneity of the original Langevin model induces the non-linear memory structure of the studied process, which is visible in the simulated data.

The work is structured as follows: in section 2 we  provide more information on the process of Brownian yet
non-Gaussian diffusion along with a primer to autoregressive models. Section 3
then introduces an intuitive physical derivation of the autoregressive model.
The situation when the correlation times of the random diffusion coefficients
is comparatively short is then considered in detail in section 4.  In section 5 we show how the statistical methods of time series can be used to qualify and quantify the non-Gaussianity. There, we also use the analytical formulas for moments to detect the non-linear memory structure specific to the considered model. We summarise our
results and discuss their utility in section 6. The paper closes with an appendix, in which the derivation of the moments is presented.}

\section{Physical stochastic modelling and autoregressive models}

\subsection{Brownian yet non-Gaussian diffusion}

In the original works on Brownian yet non-Gaussian diffusion the linear MSD
(\ref{msd}) was observed along with the Laplace shape (\ref{eq:LaplPDF}) of
the PDF. Other findings suggest the presence of more general stretched Gaussian
tails
\begin{equation}
\label{eq:stretchGauss}
p_X(x;t)\propto\exp\left(-\left(\f{|x|}{ct^{\alpha/2}}\right)^\delta\right),
\end{equation}
with the stretching parameter $\delta\neq2$. PDFs of this type describe diffusion
that can be normal ($\alpha=1$) or anomalous of the form (\ref{msd1}) with $\alpha
\neq1$. It was found that $\delta=1$ and $\alpha$ between 0.75 and 0.25 for tracer
diffusion in living bacteria and eukaryotic cells \cite{lampo17}. Similarly, in
simulations of crowded membranes it was shown that $\delta$ is between 1.3 and
1.6, and $\alpha$ below 1 \cite{jeon_prx}.\footnote{Gamma distributions were
found in \cite{hapca}, see the modelling in \cite{sposini18}.}

This behaviour can be explained by the fact that the considered medium is
spatially or temporally heterogeneous \cite{wang2,wangNG,diffDiffChechkin,
grebenkov,ralf_lampo,he16}. In such a system the thermal fluctuations are
still Gaussian but the observed displacements are mixtures of these Gaussian
contributions with random weights, effecting the non-Gaussian outcome. In
practice, this is commonly realised by two classes of diffusion models,
namely by so-called "superstatistics" and "diffusing diffusivity" models.

Superstatistics is a term proposed by Cohen and Beck \cite{beck2,beck3,beck4}
and stands for superposition of statistics. It refers to a model with random
parameters, that are fixed for each trajectory. This is a form of hierarchical
or multilevel modelling \cite{multilevel}, which is also close to the Bayesian
inference method. The distributions which arise this way are called compound
or mixture \cite{mixMod}. In the context of diffusion this mixture approach
can be explained in the following way: each observed trajectory either moves
in its own neighbourhood, that has distinct properties affecting the motion.
These features are supposed to not vary significantly at temporal and spatial
scales specific to the observed trajectories. If the motion is not confined,
this assumption must be viewed as a short-time approximation. Another possible
origin is the situation when the diffusion characteristics of the observed
particles vary from one to another, for instance, the radius of different
tracer beads. In this case the compound distributions are observed at long
times, as well, but this possibility can be excluded for experiments in which
the type of the diffusing object is precisely controlled.

\XXX{ The simplest type of a mixture model is Brownian diffusion with a randomised, but global
diffusion coefficient $D$. In it, the infinitesimal increments of the position process are given by}
\begin{equation}
\dd X(t)=\sqrt{D}\dd B(t),\quad
\end{equation}
where $B$ represents standard Brownian motion. If we assume that $D$ has an
exponential distribution, the emerging process is characterised by the Laplace
PDF \eqref{eq:LaplPDF}. Similarly, a stretched exponential PDF of $D$ leads to
relation \Ref{eq:stretchGauss}.\footnote{Of course, other forms may also emerge
from such an approach, for instance, Gamma \cite{sposini18} or stable
\cite{diffDiffChechkin} distributions.} The randomness of $D$ was confirmed
in many experiments \cite{hapca,lampo17,he16} and also in numerical studies
\cite{jeon_prx}. Since there is no fundamental reason to pinpoint the
diffusion coefficient as the only heterogeneous property of the system, other
parameters can also be made superstatistical. For example, a random viscosity
or memory kernel in the Langevin equation also lead to non-Gaussian PDFs and
generally non-linear MSD \cite{superstatLang}. It must be stressed that by
introducing a parameter which is random and fixed trajectory-wise, we break
not only homogeneity, but also locality and independence. By sharing the same
random parameter, the memory structure of the process becomes non-linear and
non-Gaussian \cite{superstatLang}. This dependence is strong enough to render
the resulting motion non-ergodic.

This non-ergodic property is not always desired and can be avoided when the
system parameters are allowed to be both random and time-dependent, as in the
diffusing diffusivity model. Fixed, deterministic values are replaced by a
stochastic variation of the parameter, and for this reason the resulting models
are called {\it doubly stochastic} \cite{doubleStoch}. If the introduced
parameter evolution is ergodic, the resulting model will be ergodic, but still
non-Gaussian. For example, one can consider a time-varying diffusion coefficient
leading to the diffusing diffusivity approach proposed by Chubinsky and Slater
\cite{diffDiff}.\footnote{In this sense the diffusing diffusivity approach is
similar in kind to the diffusing waiting times in correlated continuous time
random walks \cite{vincent,marcin}.} In this model small Brownian displacements
$\dd B(t)$ are assumed to be randomly rescaled by random $D(t)$, such that the
resulting motion is \XXX{described by the stochastic equation\footnote{Here and below such formulas should be understood as stochastic $L^2$ integrals and equations \cite{oeksendal}.}}
\begin{equation}
\label{eq:diffDiff}
\dd X(t)=\sqrt{D(t)}\dd B(t) \quad \text{or equivalently}\quad X(t)=\int_0^t\sqrt{D(s)}\dd B(s).
\end{equation}
Further details of the process depend on the specific choice of $D(t)$---the
analysis is often complicated. However, an important case was solved by Cox,
Ingerson and Ross \cite{CIR} who proposed a stochastic differential equation
with suitable mean-reverting property that leads to $D(t)$ with exponential
type of memory. Their work was motivated by financial applications and was
expressed in the language of option prices and volatility modelling. More
recently, this approach was applied to physically relevant quantities in
\cite{diffDiffChechkin,grebenkov,diffDiffCherstvy,diffDiffJain,cherayil},
in particular, for non-stationary initial conditions \cite{sposini18}.

\subsection{Autoregressive models}\label{sec:ARmod}

In contrast to models originating from physics, autoregressive models have
their roots in filtering, optimal control theory, and economics \cite{tsay00}.
In this context the measured sequence of observations $V_k$ is assumed to be
an output of a system, which is linked linearly to the white Gaussian noise
input sequence $Z_k$. The classical autoregressive moving average (ARMA)
processes  \cite{BJ,BD} are defined by the recursive relation
\begin{equation}
\label{eq:ARMA}
\fl V_k-\phi^1V_{k-1}+\phi^2V_{k-2}+\ldots\phi^pV_{k-p}=Z_k+\theta^1Z_{k-1}
+\theta^2Z_{k-2}+\ldots+\theta^qZ_{k-q}.
\end{equation}
This process is denoted by ARMA$(p,q)$ and thus explicitly contains the range
parameters $p$ and $q$. The term autoregressive AR($p$) reflects the linear
dependence of the observed $V_k$ on $p$ past values $V_{k-1},\ldots,V_{k-p}$,
and the moving average MA($q$) represents a linear combination of the last
value of the noise $Z_k$ and $q$ previous values $Z_{k-1},\ldots,Z_{k-q}$.
The admissible coefficients $\phi^i$ and $\theta^j$ are chosen such that the
sequence $V_k$ is stationary, resembling the physical velocity process (hence
the choice of notation "$V$") or highly confined motion.\footnote{For
non-stationary processes the ARIMA model ("I" stands for "integrated") may be
used: for ARIMA($p,1,q$) the differences $\Delta V_k=V_k-V_{k-1}$ are assumed
to be ARMA$(p,q)$, for ARIMA($p,n,q$) the $n$th differences are ARMA$(p,q)$.}

In most of the literature the model \Ref{eq:ARMA} was not meant to explain
the behaviour of the system. Rather, the philosophy was concentrated on
controlling the data. This can be achieved by finding a reasonably small set of
$\phi^i$ and $\theta^j$ that sufficiently well describe the observed memory
structure in the data. The procedure typically employs methods based
on mean-square optimisation and information theory \cite{BJ,BD}. Given the
estimates of the $\phi^i$ and $\theta^j$ the future behaviour can be predicted
(in the sense of the least square error or similar), which is a crucial element,
for instance, in financial forecasting. Additionally, by inverting relation
\Ref{eq:ARMA}, the $Z_k$ can be estimated from the $V_k$, therefore various
white noise tests can be used to verify the goodness of fit of the model
\cite{BD}.

It is in fact remarkable how many concepts of non-equilibrium and
biological physics have their independent counterparts in time series analysis.
For example, the study of anomalous diffusion and long-range dependence can be
performed by using autoregressive fractionally integrated (ARFIMA) models where
the white noise sequence $Z_k$ is replaced by \XXX{the fractional noise $Z_k^d$ \cite{beran, Burnecki_2014} with a power-law covariance
structure determined by the parameter $d$, $\mathbb{E}[Z_k^dZ_{k+j}^d]\propto j^{2d-1}$ (one can think of $d$ as Hurst exponent minus 1/2). It can be explicitly defined as the result of the power-law filter applied to $Z_k$,}
\begin{equation}
Z_k^d = \sum_{j=1}^\infty \f{\Gamma(j+d)}{\Gamma(d)\Gamma(j+1)} Z_{k-j}.
\label{eq:fi}
\end{equation}
For ARMA processes, the left hand side of \Ref{eq:ARMA}---the AR part--- is
responsible for modelling an exponential decay of the covariance, whereas
the MA part introduces short, finite time corrections. Additional filtering
of the white noise presented in \Ref{eq:fi} extends the ARMA model to the
ARFIMA with power-law memory, which was very successfully applied to various
anomalous diffusion phenomena encoded in biological data \cite{burnUnif}. An analogue
of L{\'e}vy flights and diffusion with a power-law PDF is the ARMA model with
stable noise, where the Gaussian $Z_k$ are replaced by non-Gaussian stable
random variables \cite{kokoszka1}. Similarly, fractional L{\'evy} stable
motion corresponds to ARFIMA with stable noise \cite{kokoszka2,Burnecki_2014}.

Another line of development in time series analysis, originating from Box-Jenkins
models, are non-linear generalisations, mainly the so-called autoregressive
conditional heteroscedasticity (ARCH) \cite{Engle_82} and generalised ARCH (GARCH)
models \cite{Boll_86}.\footnote{A set of random variables is heteroscedastic---as
opposed to homoscedastic---if there is at least one sub-population that has
different variability from the rest, where "variability" could be measured in terms
of the variance or other dispersion measures.}
These allow for the parameterisation and prediction of a
non-constant variance and proved to be very useful for financial time series; their
invention prompted the bestowal of the Nobel Memorial Prize in Economic Sciences
in 2003 to Granger and Engle. When integrated ARCH-type processes exhibit
non-Gaussian distributions for short times along with a linear time dependence
of the MSD. Precisely this observation leads us to pursue the question whether
there is a fundamental connection between the two worlds of time series analysis
using autoregressive methods and physical models of Brownian yet non-Gaussian
type.

The aforementioned notion of heteroscedasticity is related to the time-dependence
of the conditional variance \cite{breusch,nelson}. The basic ARCH$(q$) model is
defined as
\begin{equation}
\label{eq:ARCH}
V_k=\Sigma_kZ_k,\quad \Sigma_k^2=\alpha_0+\alpha_1V_{k-1}^2+\alpha_2V_{k-2}^2+
\ldots+\alpha_q V_{k-q}^2.
\end{equation}
If an ARMA model is assumed for the noise variance, the model becomes a
GARCH$(p,q)$ model,
\begin{equation}
\label{eq:GARCH}
V_k=\Sigma_kZ_k,\quad\Sigma_k^2=\alpha_0+\sum_{i=1}^q\alpha_iV_{k-i}^2+\sum_{j=1}^p
\beta_j\Sigma_{k-j}^2.
\end{equation}
A GARCH model has an ARMA-type representation, so that many of its properties
are similar to those of ARMA models, for instance, we can estimate the GARCH
parameters by the same technique as for ARMA processes \cite{tsay00}. In most
applications it is enough to consider the order of the model as $(1, 1)$.
ARFIMA combined with GARCH describes both power-law decay of the correlation
function with finite-lag effects (ARFIMA part) and varying diffusion parameter
(GARCH part) \cite{Bai_96}. For example, it can describe inhomogeneous
diffusion in the cell membrane \cite{Balcerek_19} or solar X-ray variability
\cite{Stan_19}. While the resemblance to diffusing diffusivity models
\Ref{eq:diffDiff} is indisputable: $V_k$ models velocity and $\Sigma_k$ models
the evolving diffusion coefficient, there is to date no physical derivation of
relations \Ref{eq:ARCH} and \Ref{eq:GARCH}.

Another approach related to heteroscedasticity, non-Gaussianity and linear
MSD is the main point of interest in this work: we assume that the linear
coefficients $\phi^i$ and $\theta^j$ in the definition of ARMA \Ref{eq:ARMA}
are replaced by random variables $\Phi_k^i$ and $\Theta_k^j$ that are independent
of the noise $Z_k$:
\begin{equation}
\label{eq:rcARMA}
V_k-\Phi_k^1V_{k-1}+\Phi_k^2V_{k-2}+\ldots\Phi_k^pV_{k-p}=\Theta_k^0Z_k+\Theta_k^1
Z_{k-1}+\Theta_k^2Z_{k-2}+\ldots+\Theta_k^qZ_{k-q}.
\end{equation}
This is the class of doubly stochastic models, random coefficient ARMA (rcARMA)
\cite{nicholls2012random,Roy_11}.

\section{Physical derivation of the autoregressive model}
\label{s:physDer}

Let us start with the classical Langevin equation for the velocity of a
diffusing particle,
\begin{equation}
\label{lang}
m\df{V(t)}{t}=-\beta V(t)+F(t),
\end{equation}
which is Newton's second law with the Stokes dissipative force $-\beta V$ and
the stochastic force $F$. In the classical setting $F$ is given by white Gaussian
noise with amplitude $k_BT\beta$ determined by the fluctuation-dissipation
relation \cite{zwanzig,coffey}. Heterogeneity of the medium can be modelled
by making the parameters of equation (\ref{lang}) time-dependent and random---then
they describe the local state of the environment of the diffusing particle. All
possible correlation effects caused by the particle repeatedly visiting the same
areas of the phase space are assumed to be reflected by the memory structure of
random functions modelling these local parameters. Starting from the next section
we will neglect this dependence entirely, which corresponds to an annealed picture
that is also inherent in the current physical diffusing diffusivity models.This
assumption can be also justified when the environment changes continuously. The
resulting equation can then be written in the form 
\begin{equation}
\label{eq:timeDepLang}
\dd V(t)=-\Lambda(t)V(t)\dd t+\sqrt{D(t)}\dd B(t).
\end{equation}
We do not provide any more fundamental derivation behind this formula, for us
$\Lambda(t)$ and $D(t)$ are effective parameters whose physical meaning is
determined by equation \Ref{eq:timeDepLang} alone: $D(t)$ describes the local
effective amplitude of velocity gains and $\Lambda(t)$ the local effective
linear damping or relaxation of the velocity. In financial language these
parameters would be called stochastic return rate and stochastic volatility. \XXX{ Take note this class of models differs from the superstatistical approach of Beck and Cohen \cite{beck2}, who assume the Langevin dynamics and stochasticity of environement can be considered separately. In \Ref{eq:longPeriodS} we consider a particular case leading to superstatistics.  }

Formula \Ref{eq:timeDepLang} resembles the rcARMA. Indeed, in the standard
Euler scheme of numerical simulations, one fixes a time delay $\Delta t$
between observations, denotes $V_k=V(k\Delta t)$ and approximates $\dd
V(t)\approx V_{k}-V_{k-1}$, which results in
\begin{equation}
V_k-(1-\Lambda_k\Delta t)V_{k-1}=\sqrt{D_k}\Delta B_k.
\end{equation}
This is an rcAR(1) process with AR coefficient $1-\Lambda_k\Delta t$. It is
easy to see that the same reasoning applies to any linear stochastic
differential equation and thus the classical schemes of discrete-time
simulations return rcARMA processes approximating the continuous-time
solutions.

In fact, the connections between rcARMA and models of diffusing diffusivity
reach deeper than that. Classical results from the theory of time series
analysis establish that there is an exact correspondence between ARMA and
solutions of linear stochastic equations with constant coefficients---no
approximation is needed \cite{BrockwellCAR}. The classical Langevin equation
for a position of a particle confined in an harmonic potential,
\begin{equation}
\label{eq:hp}
m\frac{\dd^2 X(t)}{\dd t^2}=-\kappa X(t)-\beta\frac{\dd X(t)}{\dd t}+F(t)
\end{equation}
fits into this category. The resulting discrete motion is the ARMA(2,1) process
\begin{equation}
\label{eq:hp2}
X_k-\phi^1X_{k-1}-\phi^2X_{k-2}=Z_k+\theta Z_{k-1}
\end{equation}
with AR(2) coefficients
\begin{align}
\phi^1&=\e^{-\Delta t\frac{\beta}{2m}}\left(\e^{\Delta t\sqrt{\left(\frac{
\beta}{2m}\right)^2-\frac{\kappa}{m}}}+\e^{-\Delta t\sqrt{\left(\frac{\beta}{2m}
\right)^2-\frac{\kappa}{m}}}\right),\nonumber\\ \phi^2&=-\e^{-\Delta t\frac{
\beta}{m}}.
\end{align}
The MA(1) coefficient $\theta$ can be easily calculated numerically or expressed
by a somewhat complicated but elementary formula \cite{slezakWeron}. This class
of relations has direct application to the statistical analysis of the data and
can be used, for instance, to calculate the exact discrete-time power spectral
density (the spectrum given by the Fourier series of $X_k$) without the need of
the commonly used approximation $\sum_k f(X_k)\Delta t\approx\int f(X(t))\dd t$
\cite{drobczynskiSlezak}.

\begin{figure}
\centering
\includegraphics[width=10cm]{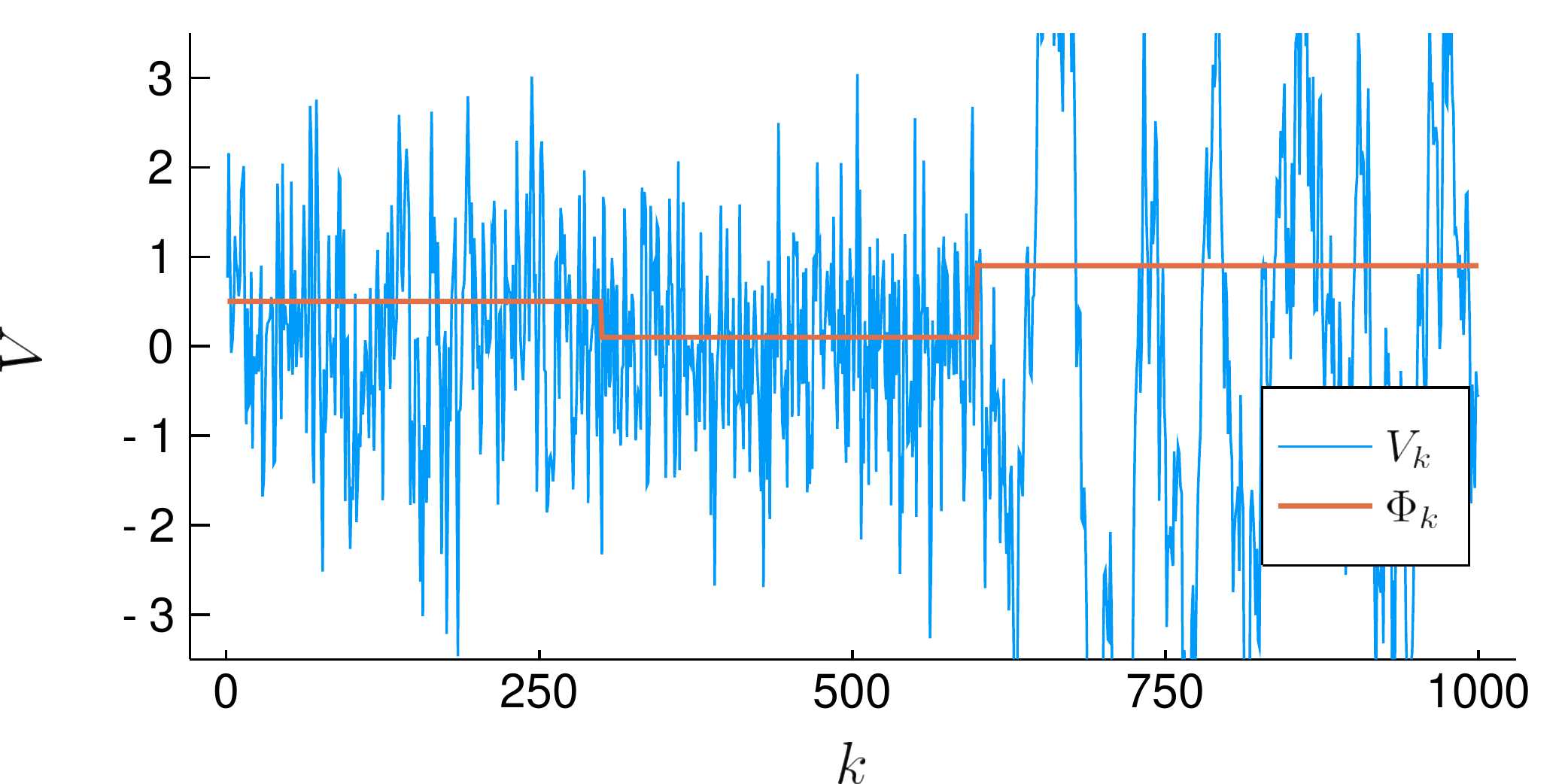}
\caption{Trajectory $V_k$ in which $\Phi_k$ changes in 3 intervals from $0.5$,
to $0.1$, and finally $0.9$, while $\Theta_k=1$ is kept constant. For small
$\Phi_k$ the motion resembles white noise, larger values result in longer
Brownian excursions away from $0$. Note that for better visibility very long
excursion are cut off.}
\label{fig:traj}
\end{figure}

The core of the physical interpretation we propose in this work is that the
derivation used to obtain \Ref{eq:hp} and \Ref{eq:hp2} can be generalised
for the case of linear equations with time-dependent coefficients. Namely, we
multiply \Ref{eq:timeDepLang} by the integrating factor $\exp(\int\Lambda(t)
\dd t)$ and integrate from $(k-1)\Delta t$ to $k\Delta t$, obtaining
\begin{equation}
\e^{\int_0^{k\Delta t}\Lambda(s)\dd s}V_k-\e^{\int_0^{(k-1)\Delta t}\Lambda(s)\dd
s}V_{k-1}=\int_{(k-1)\Delta t}^{k\Delta t}\e^{\int_0^{s} \Lambda(s')\dd s'} \sqrt{
D(s)}\dd B(s).
\end{equation}
After some rearrangement it takes the sleek form
\begin{equation}
V_k-\Phi_kV_{k-1}=Z_k
\end{equation}
with 
\begin{equation}
\label{eq:PhiZDef}
\Phi_k\defeq \e^{-\int_{(k-1)\Delta t}^{k\Delta t}\Lambda(s)\dd s},\quad
Z_k\defeq \int_{(k-1)\Delta t}^{k\Delta t}\e^{-\int_s^{k\Delta t}\Lambda(s')\dd s'}
\sqrt{D(s)}\dd B(s).
\end{equation}
The random coefficient AR form of the left hand side is clearly present, but it is
not immediately clear what is the structure of $Z_k$ on the right. Different
$Z_k$ use increments $\dd B(s)$ from disjoint intervals $[(k-1)\Delta t,k\Delta
t]$. The Gaussian increments $\dd B(s)$ are stationary and do not depend on $k$
but they are rescaled by $D(t)$ and the exponent of $\Lambda(t)$, which makes
their conditional variances a sequence of random variables,
\begin{align}
\mathbb{E}[Z_k^2|\Lambda,D]&=\int_{(k-1)\Delta t}^{k\Delta t}\left(
\e^{-\int_{s}^{k\Delta t}\Lambda(s')\dd s'}\sqrt{D(s)}\right)^2\dd s\nonumber\\
&=\int_{(k-1)\Delta t}^{k\Delta t}\e^{-2\int_s^{k\Delta t}\Lambda(s')\dd s'}
D(s)\dd s.
\end{align}
Conditioned on $\Lambda$ and $D$ the variables $Z_k$ are Gaussian, independent,
and have the variance specified above. Thus the $Z_k$ can be decomposed as $Z_k
=\Theta_k W_k$ for a Gaussian white noise $W_k$ and coefficients $\Theta_k=\sqrt{
\mathbb{E}\left[Z_k^2|\Lambda,D\right]}$, which is a function of random $\Lambda
(t)$ and $D(t)$. This is but exactly the rcARMA
\begin{equation}
\label{eq:rcAR}
V_k-\Phi_k V_{k-1}=\Theta_kW_k,
\end{equation}
in which the regressive coefficients $0\le\Phi_k\le1$ model the local return
or relaxation rates, and the amplitude coefficients $\Theta_k\ge0$ describe the heteroscedasticity
of fluctuations, that is the variability of the fluctuations' dispersion. An
illustration how the changes of $\Phi_k$ are visible in the data can be seen
in figure \ref{fig:traj}.

Each $\Phi_k$ is a decreasing functional of the values $\Lambda(t)$ from the
interval $[(k-1)\Delta t,k\Delta t]$. Similarly $\Theta_k$ is increasing with
$D(t)$ and decreasing with $\Lambda(t)$. Therefore $\Phi_k$ and $\Theta_k$
are two sequences which should mirror the memory structure of $\Lambda(t)$
and $D(t)$. They are dependent, because they both are functionals of the
same range of values $\Lambda(t)$. In general the form of this dependence is
highly non-linear. When the parameters can be assumed to not vary significantly
in the interval $\Delta t$, $\Lambda(t)\approx\Lambda_k$ and $D(t)\approx D_k$,
the approximate relation reads
\begin{equation}
\Phi_k\approx\e^{-\Delta t\Lambda_k},\quad\Theta_k\approx\sqrt{\f{D_k}{2\Lambda_k}}
\sqrt{1-\e^{-2\Delta t\Lambda_k}}\approx\sqrt{\f{\Delta tD_k}{-2\ln(\Phi_k)}}\sqrt{
1-\Phi_k^2}.
\end{equation}
In practical applications this formula can be further simplified. Namely, the
function $(1-x^2)/(-\ln(x))$ is approximately linear on the interval $[0,1]$,
thus a reasonable approximation is $\Theta_k\approx\sqrt{D_k\Phi_k}$.

No matter what is the distribution of the $\Phi_k$ and $\Theta_k$, the solution
of \Ref{eq:rcAR} itself can be expressed in a simple manner. Repeating the
recursive relation between $V_k$ and $V_{k-1}$ we can write the explicit
formula linking $V_k$ and $V_{k-n}$ for any $n$ in the form
\begin{align}
\label{eq:recV}
V_k&=\Phi_k\Phi_{k-1}\cdots \Phi_{k-n+1}V_{k-n}+\Theta_kW_k
+\Phi_k\Theta_{k-1}W_{k-1}+\Phi_k\Phi_{k-1}\Theta_{k-2}W_{k-2}+\ldots\nonumber\\
&+\Phi_k\cdots\Phi_{k-n+2}\Theta_{k-n+1}W_{k-n+1}.
\end{align}
Away from the degenerate case $\Lambda(t)\equiv0$ it is true that $\Phi_k<1$,
therefore when $n\to\infty$ the series converges towards
\begin{equation}
\label{eq:seriesRep}
V_k=\sum_{i=0}^{\infty}\Big(\prod_{j=0}^{i-1}\Phi_{k-j}\Big)\Theta_{k-i}W_{k-i}.
\end{equation}
We see that each $V_k$ is a mixture of Gaussian variables $W_k$ with random
weights. The result is not Gaussian, but can be characterised as conditionally
Gaussian $\mathcal N(0,S^2)$ with conditional variance
\begin{equation}
\label{eq:S2}
S^2\defeq\mathbb{E}\left[V_k^2|\Phi,\Theta\right]=\sum_{i=0}^{\infty}\Big(
\prod_{j=0}^{i-1}\Phi_{k-j}^2\Big)\Theta_{k-i}^2.
\end{equation}
The distribution of $S^2$ determines the non-Gaussianity: the more spread out
$S^2$ is, the more the PDFs of the velocity and displacement differ from the
Gaussian shape. This property can be expressed in terms of the excess kurtosis,
which is directly related to the relative standard deviation (RSD) of $S^2$.
Namely,
\begin{equation}
\label{eq:kurt}
\kappa_V \defeq\f{\mathbb{E}\left[V_k^4\right]}{\mathbb{E}\left[V_k^2\right]^2}-3
=3\f{\mathbb{E}\left[(S^2-\mathbb{E}\left[S^2\right])^2\right]}{\mathbb{E}\left[
S^2\right]^2}
\end{equation}
is three times the RSD squared. Different variants of the excess kurtosis or the
RSD of $S^2$ are sometimes called "non-Gaussianity parameter" \cite{matthias1,
lanoiselee}, but here we will use more precise terminology. Because both are positive,
the distribution of $V$ is always leptokurtic, that is, it has tails thicker than
a Gaussian. It is worth stressing that the kurtosis is only one of many possible
measures of non-Gaussianity, but it is a convenient one because of the easy
estimation.

\begin{figure}
\centering
\includegraphics[width=10cm]{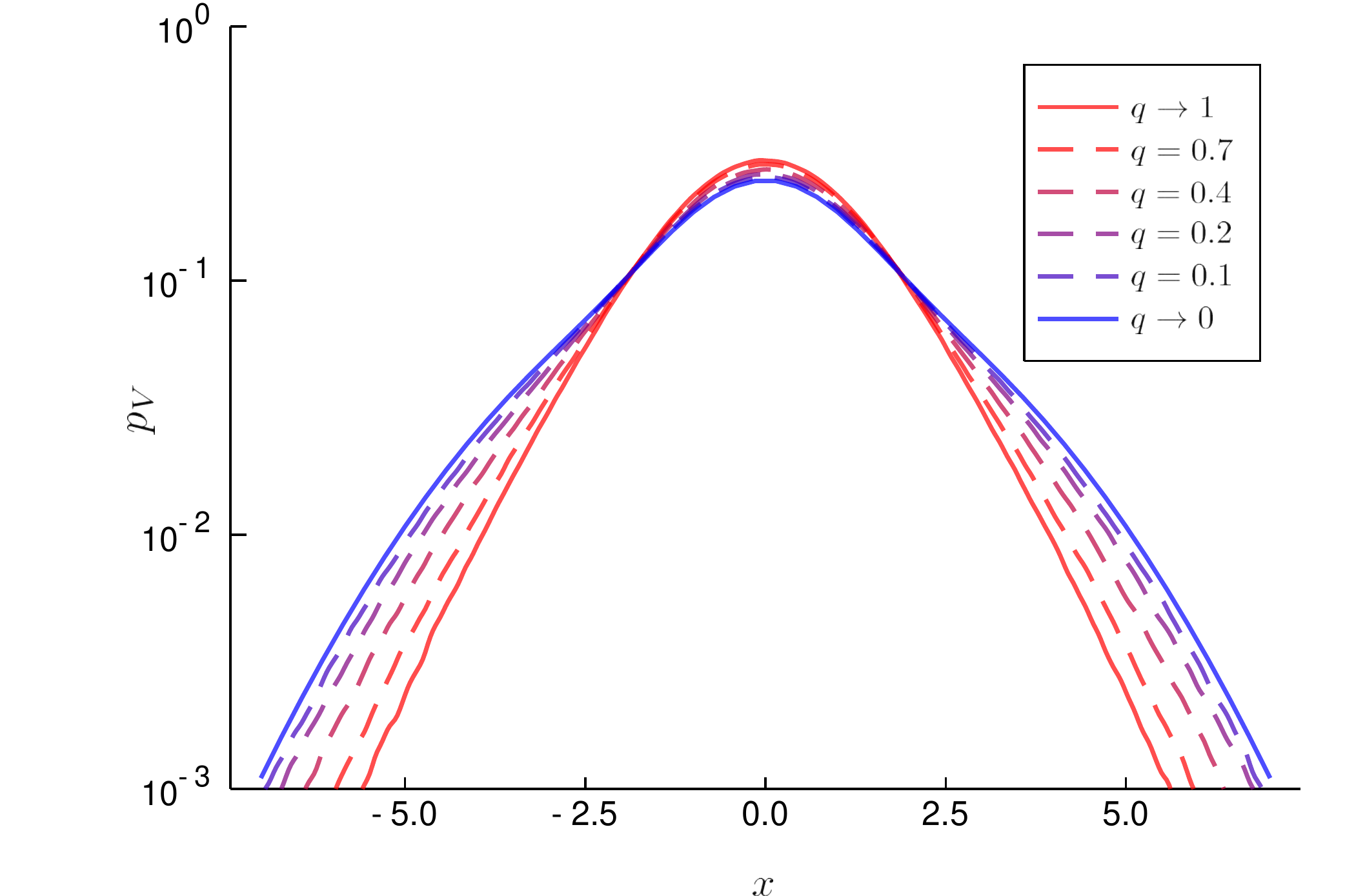}
\caption{PDF of the rcAR process $V$ with random periods $T_\alpha$ of constant
coefficients $\Phi_\alpha$ and $\Theta_\alpha=1$. The dashed lines represent rcAR
simulations, the solid lines correspond to theoretical predictions for the limiting
cases. The distribution was chosen to be the simple \XXX{geometric distribution $\mathcal{G}eo(q))$, $\mathrm{P}(\mathcal{G}eo(q)) = k) = q^k$, and the two types of periods are chosen as $\mathrm{P}(
\Phi_\alpha=0.1,T_\alpha\deq\mathcal{G}eo(q))=1/2$ and $\mathrm{P}(\Phi_\alpha=0.9,T_
\alpha\deq \mathcal{G}eo(q))+\mathcal{G}eo(q))=1/2$.} The periods of $\Phi_\alpha
=0.9$ are on average twice longer, thus the effective distribution is $\mathrm{P}
(\widetilde\Phi=0.1)=1/3$, and $\mathrm{P}(\widetilde\Phi=0.9)=2/3$. The prediction
$q\to1$ from \Ref{eq:S2} with independent coefficients distributed like $\widetilde\Phi$.
The prediction $q\to0$ comes from \Ref{eq:longPeriodS}, in the latter case the PDF
is just a superposition of two Gaussians.}
\label{fig:intPDF}
\end{figure}

The distribution of $S^2$ is, in general, hard to analyse even for simple models
of $\Phi$ and $\Theta$. One exception is encountered when the coefficients stay
constant for long intervals of time $T_\alpha$ (precisely, $\mathbb{E}\left[T_
\alpha\right]$ larger than the mean relaxation time $1/\mathbb{E}\left[\Phi
\right]$). In such a situation $S^2$ spends short times transitioning between
these periods and therefore in the majority they attain an equilibrium value
obtained from taking the coefficients in \Ref{eq:S2} constant,
\begin{equation}
S^2_\alpha=\f{\Theta^2_\alpha}{1-\Phi_\alpha^2},
\end{equation}
where $\alpha$ is an identification of the interval. The resulting distribution
of $S^2$ is thus
\begin{equation}
\label{eq:longPeriodS}
S^2\approx\f{\widetilde\Theta^2}{1-\widetilde\Phi^2},
\end{equation}
with $\widetilde\Theta$ and $\widetilde\Phi$ being random variables constructed
from the values $\Theta_\alpha,\Phi_\alpha,\Theta_\beta,\Phi_\beta,\Theta_\gamma,$
$\Phi_\gamma,\ldots$ with probabilities weighted by the relative mean lengths of
the intervals $T_\alpha,T_\beta,T_\gamma,...$. In this approximation one trajectory
is effectively the same as an ensemble of equal-length trajectories, each with its
own $\widetilde\Theta,\widetilde\Phi$, which are glued together at the ends, merging
into a single trajectory. \XXX{(This is essentially a form of superstatistics.)} For a simple demonstration see figure \ref{fig:intPDF}.
There one of the limiting cases are long periods of constant coefficients, as in
\Ref{eq:longPeriodS}, but one can also observe the second limit of rapidly
varying coefficients. This is the subject of the next section.

\section{Short memory random coefficient AR(1)}

\begin{figure}
\centering
\includegraphics[width=7.6cm]{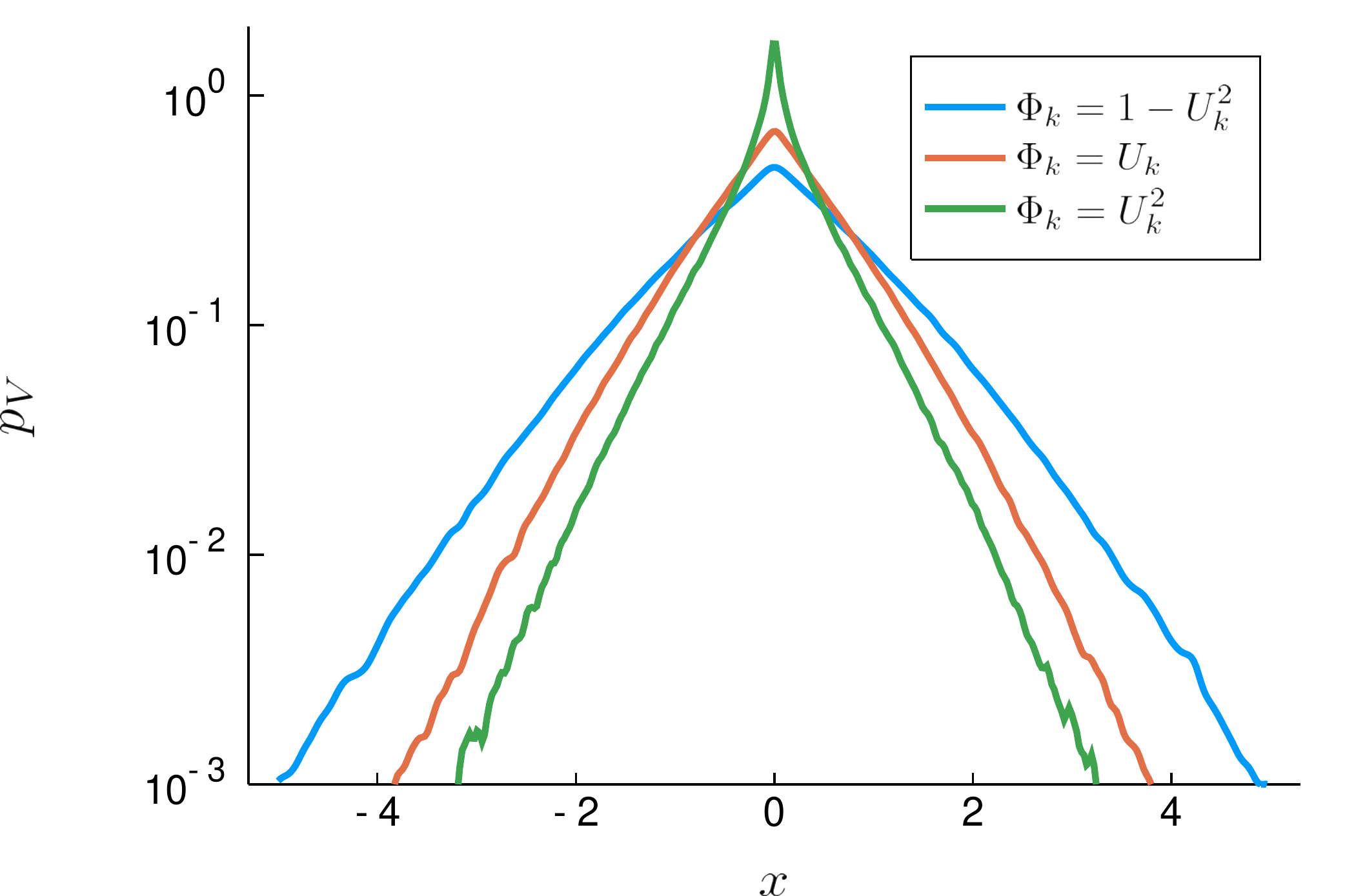}
\includegraphics[width=7.6cm]{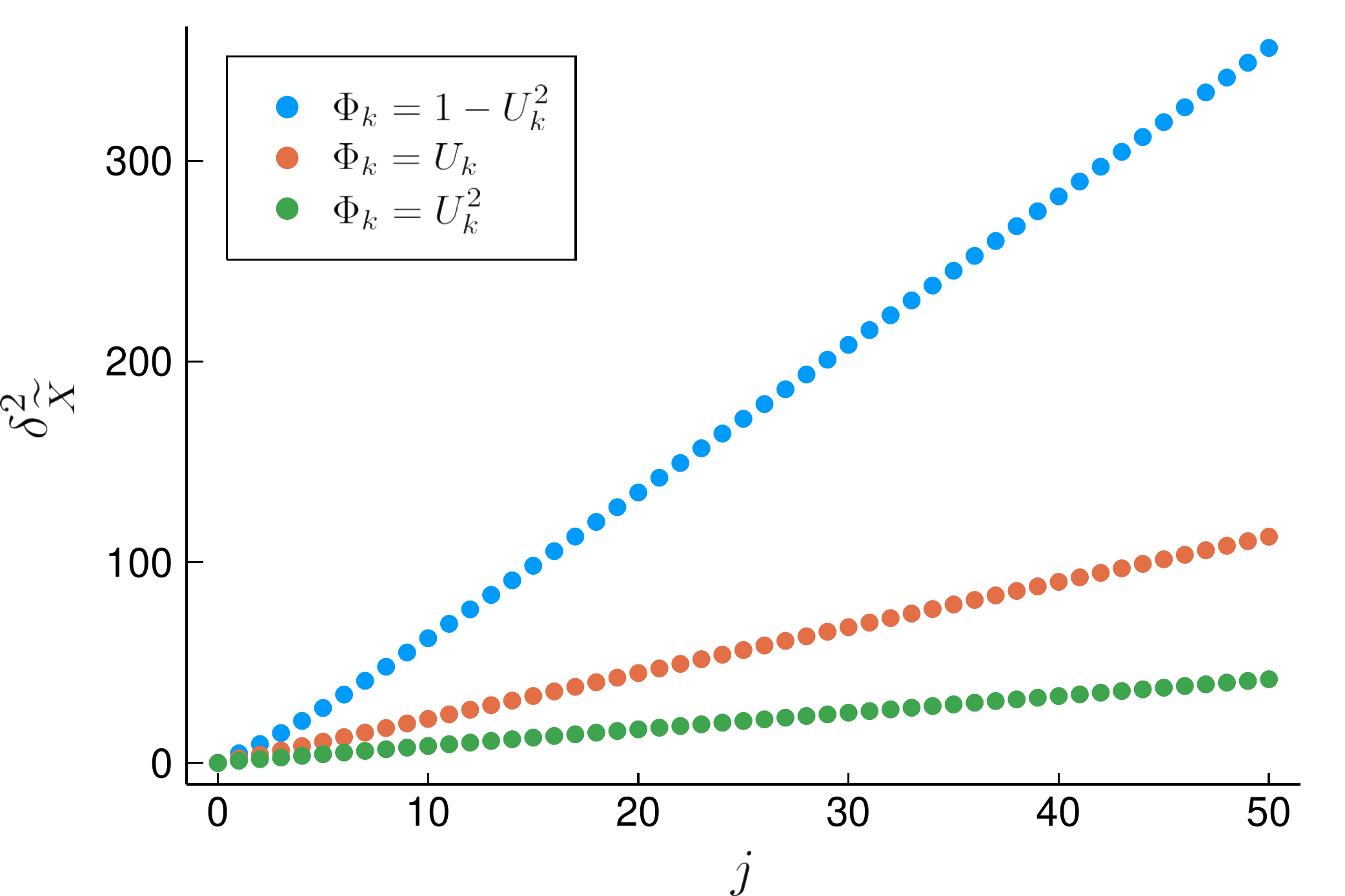}
\caption{PDF and MSD estimated from the rcAR process $V$. Here $U_k$ are independent
and uniformly distributed, $\mathcal U(0,1)$, and $\Theta_k=\sqrt{\Phi_k}$.
Different transformations of $U_k$ correspond to different power laws of
$\Phi_k$ around $0^+$ and $1^-$.}
\label{fig:pdfComp}
\end{figure}

When the correlation time of $\Lambda(t)$ and $D(t)$ is much smaller than the
observation time, $t_c\ll\Delta t$, the integrals defining $\Phi_k,\Phi_{k+1}$
and $\Theta_k,\Theta_{k+1}$ contain only the ratio $t_c/\Delta t$ of the total
mass corresponding to dependent values of $\Lambda(t),D(t)$. Thus, $\Phi_k$
and $\Theta_k$ can be assumed to be sequences of independent and identically distributed (i.i.d.) variables. Note
that for fixed $k$ the pair $\Phi_k,\Theta_k$ may, and generally will be dependent. Averaging
\Ref{eq:S2} yields
\begin{equation}
\label{eq:varV}
\mathbb{E}\left[V_k^2\right]=\mathbb{E}\left[S^2\right]=\f{\mathbb{E}\left[
\Theta_k^2\right]}{1-\mathbb{E}\left[\Phi_k^2\right]}.
\end{equation}
Similarly, $\mathbb{E}\left[S^4\right]$ can be calculated, as detailed in the
appendix. The moments $\mathbb{E}\left[S^2\right]$ and $\mathbb{E}\left[S^4
\right]$ together determine the kurtosis of $V$. Accordingly, the motion is
again non-Gaussian. The covariance also has a simple form: from the recursive
formula \Ref{eq:recV} we can easily derive the geometric decay
\begin{equation}
\label{eq:rcARcov}
r_V(j)\defeq \mathbb{E}\left[V_kV_{k+j}\right]=\mathbb{E}\left[S^2\right]\varphi^j,
\quad \varphi\defeq\mathbb{E}\left[\Phi_k\right].
\end{equation} 
The form of the covariance also determines the MSD: for the velocity it converges
exponentially to a constant $\delta^2_V(j)=2(1-r_V(j))$ and for the position $X_k
\defeq\int_0^{k\Delta t}V(\tau)\dd\tau $ it is linear for $j\gg 1$,
\begin{equation}
\delta_X^2(j)=\f{c^2}{-\ln\varphi}j+\f{c^2}{\ln\varphi^2}\left(\varphi^j-1\right),
\quad c^2\defeq 2\Delta t^2\mathbb{E}\left[S^2\right].
\end{equation}
A similar formula can be obtained if one tries to recreate displacements given
$V_k$: then $\widetilde X_k\defeq\Delta t\sum_{i=1}^kV_i$ leads to the formula
\begin{equation}
\delta_{\widetilde X}^2(j)=\f{c^2}{2}\f{(1+\varphi)}{1-\varphi}j-c^2\varphi\f{1
-\varphi^j}{(1-\varphi)^2}.
\end{equation}
A small error in determining the slope of the true $X(t)$ is made in this
procedure, but the general behaviour of the MSD is the same. Note that the
shapes of all the above functions depend only on the mean value of $\Phi_k$,
this is a diffusion with covariance and MSD (up to a scaling factor) as if
the system were homogenised---however, at the same time it is non-Gaussian,
see figure \ref{fig:pdfComp}.

\begin{figure}
\centering
\includegraphics[width=10cm]{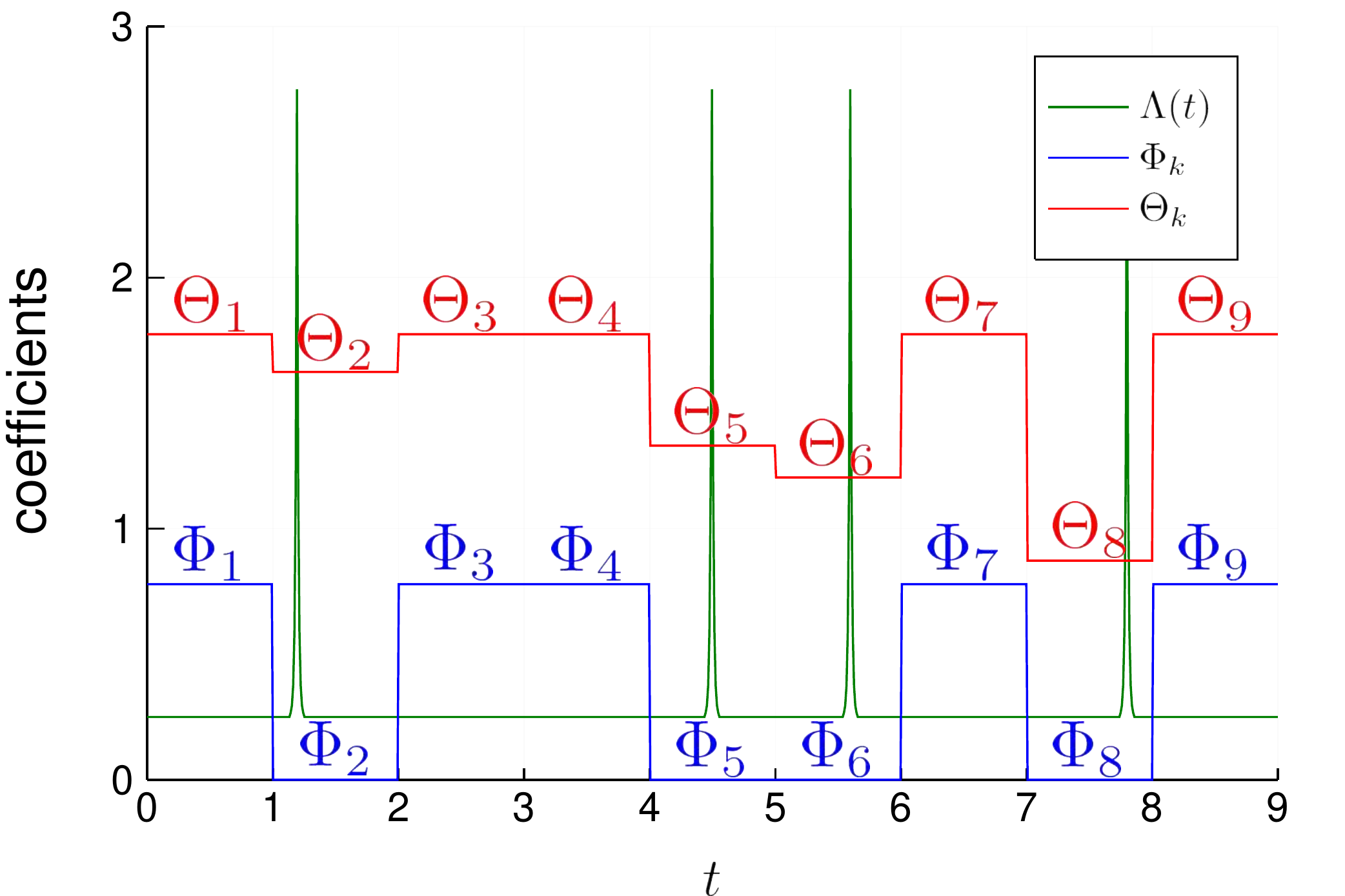}
\caption{Damping rate $\Lambda(t)$ and corresponding $\Phi_k,\Theta_k$. Here
$\Delta t=1$ and, outside of the short impulses, $\Lambda(t)=1/4$. To better
highlight different values of the coefficients, the diffusion coefficient was
taken to be $D(t)=4$.}
\label{fig:coeff}
\end{figure}

As a concrete example of a physical model which leads to this type of process
let us  assume that the diffusivity is constant (with no loss of generality we
take $D(t)=1$). Then we consider a diffusing particle interacting with high
viscosity traps distributed in a Poissonian manner, that is, the waiting time
between subsequent trapping events have the exponential distribution $\mathcal
E(\rho)$. By a high-viscosity trap we mean a short-time interval when $\Lambda
(t)=\infty$, which immediately kills the momentum of the particle---this is a
form of stochastic resetting \cite{BrownResett}. Outside of these events the
damping rate is constant $\Lambda(t)=\lambda$. The probability that the particle
does not meet any traps between times $(k-1)\Delta t$ and $k\Delta t$ is $\exp(
-\rho\Delta t)$: in this case $\Phi_k=\exp(-\lambda\Delta t)$ and $\Theta_k=
\sqrt{(1-\exp(-2\lambda\Delta t))/(2\lambda)}$. The probability that the
particle meets at least one trap in an interval $[(k-1)\Delta t,k\Delta t]$ is
of probability $1-\exp(-\rho\Delta t)$, and the corresponding AR(1) coefficient
$\Phi_k$ becomes 0. The discrete noise $Z_k$ for this $k$ has a more complex
structure. All Gaussian fluctuations before the last trapping event are erased,
which is visible from \Ref{eq:PhiZDef}. However the remaining fluctuations still
count. Denoting by $T_k$ the difference between the last trapping event and $k
\Delta t$ we obtain the formula for $\Theta_k'$ corresponding to the event
$\Phi_k=0$,
\begin{equation}
\label{eq:impulseTheta}
\Theta_k'=\sqrt{\int_{k\Delta t-T_k}^{k\Delta t}\e^{-2\lambda(k\Delta
t-s)}\dd s}=\sqrt{\f{1-\e^{-2\lambda T_k}}{2\lambda}}\approx \sqrt{T_k}.
\end{equation}
This corresponds to a small correction $\mathcal O(\Delta t)$, which makes the
resulting PDF smooth (otherwise the trapping events would be visible as a Dirac
delta at $x=0$). The approximation on the right holds for $\lambda\Delta t\ll1$.
The relation between $\Lambda(t)$ and $\Phi_k,\Theta_k$ is shown in figure
\ref{fig:coeff} using an exemplary trajectory.

Variables $T_k$ which determine $\Theta_k$ have the distribution of an exponential
variable conditioned by the requirement that it has value lower than $\Delta t$
(that is, the trapping event occurred in the considered interval). It has the PDF
\begin{equation}
p_{T_k}(t)=\f{\rho}{1-\e^{-\rho\Delta t}}\e^{-\rho t},\quad 0\le t\le\Delta t.
\end{equation}
As we determined the exact distribution of $\Phi_k$ and $\Theta_k$, the formula
for the covariance \Ref{eq:rcARcov} can be made completely explicit, namely
\begin{equation}
r_V(j)=\mathbb{E}\left[S^2\right]\e^{-j(\rho+\lambda)\Delta t},\quad
\mathbb{E}\left[S^2\right]=\f{1}{2\lambda}\f{1+\e^{-\rho\Delta t}-2\f{\rho+
\lambda}{\rho+2\lambda}\e^{-(\rho+2\lambda)\Delta t}}{1-\e^{-(\rho+2\lambda)
\Delta t}}.
\end{equation}
The variance $\mathbb{E}\left[S^2\right]$ has a complicated form, but the
geometric decay rate is given by $(\rho+\lambda)\Delta t$. It is an intuitive
result: meeting a trap erases the momentum of the particle, which thus loses
all connection to its history at a frequency proportional to the density of
traps.

Studying the distribution of $V_k$ in general is hard, even for i.i.d.
coefficients, because the conditional variance \Ref{eq:S2} is given by an
infinite sum with terms which are dependent, as the same $\Phi_{k-i}$
reappear in them. In the studied example the situation is specific and
simpler, because the sum is actually finite. The coefficients $\Phi_{k-i}$
are a series of Bernoulli trials---after a series of $j-1$ non-zero ones
appearing with probability $\exp(-\rho \Delta t)^j$ the first zero occurs
with probability $1-\exp(-\rho\Delta t)$ and the summation stops. Conditioned
by this event the first $j-1$ values $\Theta_{k-i}$ are deterministic, but the
last term in the sum, $\Theta_{k-j}'$ has the conditional distribution
\Ref{eq:impulseTheta}. This leads to the formula for characteristic function,
\begin{align}
\widehat p_V(\theta)&=\mathbb{E}\left[\exp\left(-\f{\theta^2}{2}S^2\right)\right]\\
\nonumber
&=\sum_{j=0}^\infty\mathbb{E}\left[\exp\left(-\f{\theta^2}{2}\left(\sum_{i=0}^{j-1}
\e^{-2i\lambda\Delta t}\Theta_{k-i}^2+\e^{-2j\lambda\Delta t}\Theta_{k-j}'^2\right)
\right)\right]\\\nonumber
&\times\e^{-j\rho\Delta t}\left(1-\e^{-\rho\Delta t}\right)\nonumber\\
&=\left(1-\e^{-\rho\Delta t}\right)\e^{-\f{\theta^2}{4\lambda}}\sum_{j=1}^\infty
\exp\left(\f{\theta^2}{4\lambda}\e^{-2j\lambda\Delta t}\right)\e^{-j\rho\Delta t}
\mathbb{E}\left[\exp\left(-\f{\theta^2}{2}\e^{-2j\lambda\Delta t}\Theta_{k-j}'^2
\right)\right].
\end{align}
Here the Laplace transform of the conditional $\Theta_{k-j}'\deq\Theta_{k-j}|\{\Phi
_{k-j}=0\}$ is given by
\begin{align}
\mathbb{E}\left[\e^{-s\Theta_{k-j}^2}|\Phi_{k-j}=0\right]&=\rho\f{\e^{-\f{s}{2
\lambda}}}{1-\e^{-\rho\Delta t}}\int_0^{\Delta t} \e^{\f{s}{2\lambda}\e^{-2\lambda
t}}\e^{-\rho t}\dd t\nonumber\\
&=\f{\rho}{2\lambda}\left(\f{2\lambda}{s}\right)^\f{\rho}{2\lambda}\f{\e^{-\f{s}{
2\lambda}}}{1-\e^{-\rho\Delta t}}(-i)^{\f{\rho}{\lambda}}\left(\Gamma\left(\f{\rho}{
2\lambda},-\f{s}{2\lambda}\right)\right.\nonumber\\
&\left.-\Gamma\left(\f{\rho}{2\lambda},-\f{s}{2\lambda}\e^{-2\lambda\Delta t}\right)
\right)\nonumber\\
&=\rho\f{\e^{-\f{s}{2\lambda}}}{1-\e^{-\rho\Delta t}}\sum_{j=0}^\infty \f{1}{j!}
\left(\f{s}{2\lambda}\right)^j\f{1-\e^{-(2j\lambda+\rho)\Delta t}}{2j\lambda+\rho}
\approx \f{\rho}{s+\rho} \f{1-\e^{-(s+\rho)\Delta t}}{1-\e^{-\rho\Delta t}},
\end{align}
The last  approximation works for $\lambda\Delta t\ll 1$ which follows from
\Ref{eq:impulseTheta}. The limiting case $\lambda \Delta t \approx 0$ is also
interesting: then $\Phi_k$ are i.i.d. variables with Bernoulli distribution
$\mathrm{P}(\Phi_k=1)=\exp(-\rho\Delta t)=1-\mathrm{P}(\Phi_k=0)$, as usual
$\Theta_k=\sqrt{\Phi_k}$---for simplicity we may neglect the fluctuations
counted in the next interval after the trapping events. Directly from
$\Ref{eq:S2}$ we see that $S^2$ has a geometric distribution. Any such
variable can be expressed as an integer part of some exponentially distributed
variable, $E$ say, and therefore it has values between $E$ and $E-1$. As
mentioned in the introduction, the mixture of Gaussian variables with variance
$E$ has exponential tails \cite{Kotz}, so this is the case for $S^2$, as well. This
also provides an intuitive argument for the presence of the tails with thickness
between Gaussian and Laplace in the general case considered in this section (see
figure \ref{fig:pdfComp}). As the $\Phi_k$ are bounded by $1$, the tails of $S^2$
correspond to long stretches of $\Phi_k$ close to $1$, and as they are i.i.d. the
probability of such an event decays geometrically with the number of $\Phi_k$.

\section{Statistical analysis}
\label{sec:stat}

We now address the statistical procedures in the context of stochastic processes
with random coefficients.

\subsection{Memory}

In \Ref{eq:rcARcov} we showed that the covariance of i.i.d. rcAR(1) is, up
to a scale factor, the same as for the AR(1) with coefficient $\phi=
\mathbb{E}\left[\Phi_k\right]$. The same is also true for general i.i.d. rcARMA,
it can be demonstrated by multiplying both sides of the equation by past
values of the process and deriving the recursive formulas for the covariance
(these are called Yule-Walker equations)---they differ from the non-random
coefficient case only by the single equation which determines the variance
\cite{liangRCA}. On the one hand it makes the additional randomness harder
to detect, but on the other hand it allows to use powerful ARMA based methods
of analysis. As we mentioned in section \ref{sec:ARmod}, AR (thus also rcAR)
have a covariance function that is a mixture of exponentials, and the MA part is
responsible for short time corrections. For the continuous-time processes the
only option of statistical verification would seem to be fitting the estimated
covariance function and checking the validity of the result. For rcARMA a more
intuitive tool is available: the partial autocorrelation function. Given a
stationary time series $X_k$ it measures the linear dependence between $X_k$
and $X_{k-j}$ when an in-between set of variables $\mathcal I_{k,j}\defeq X_
{k-1},X_{k_2},\ldots X_{k-j+1}$ is removed. Denoting the least-squares
projection operator by $\mathcal P$ it can be defined as
\begin{equation}
\alpha_X(j)\defeq\mathbb{E}\left[X_k^2\right]^{-1}\mathbb{E}\left[(X_k-
\mathcal P_{\mathcal I_{k,j}} X_k)(X_{k-j}-\mathcal P_{\mathcal I_{k,j}}
X_{k-j})\right].
\end{equation}
The partial autocorrelation can be effectively estimated using standard
methods implemented in the popular statistical packages (such as Python
\texttt{Statsmodels}, R \texttt{tseries}, Julia \texttt{StatsBase}, or Matlab
Econometrics Toolbox)---the usual command is the standard abbreviation
\texttt{pacf}. It is intuitively clear that $\alpha_X$ measures the "direct"
dependence between $X_k$ and $X_{k-j}$ which for AR(p) and rcAR(p), given
their recursive definition, should be non-zero for the first $p$ values
and zero further on. For the  MA part one can reverse the definition and
express the noise $Z_k$ as the geometrical sum of $X_k$, hence, the partial
autocorrelation decays geometrically \cite{BJ,BD}. For the full ARMA model
the two effects are additive.

\begin{figure}
\centering
\includegraphics[width=10cm]{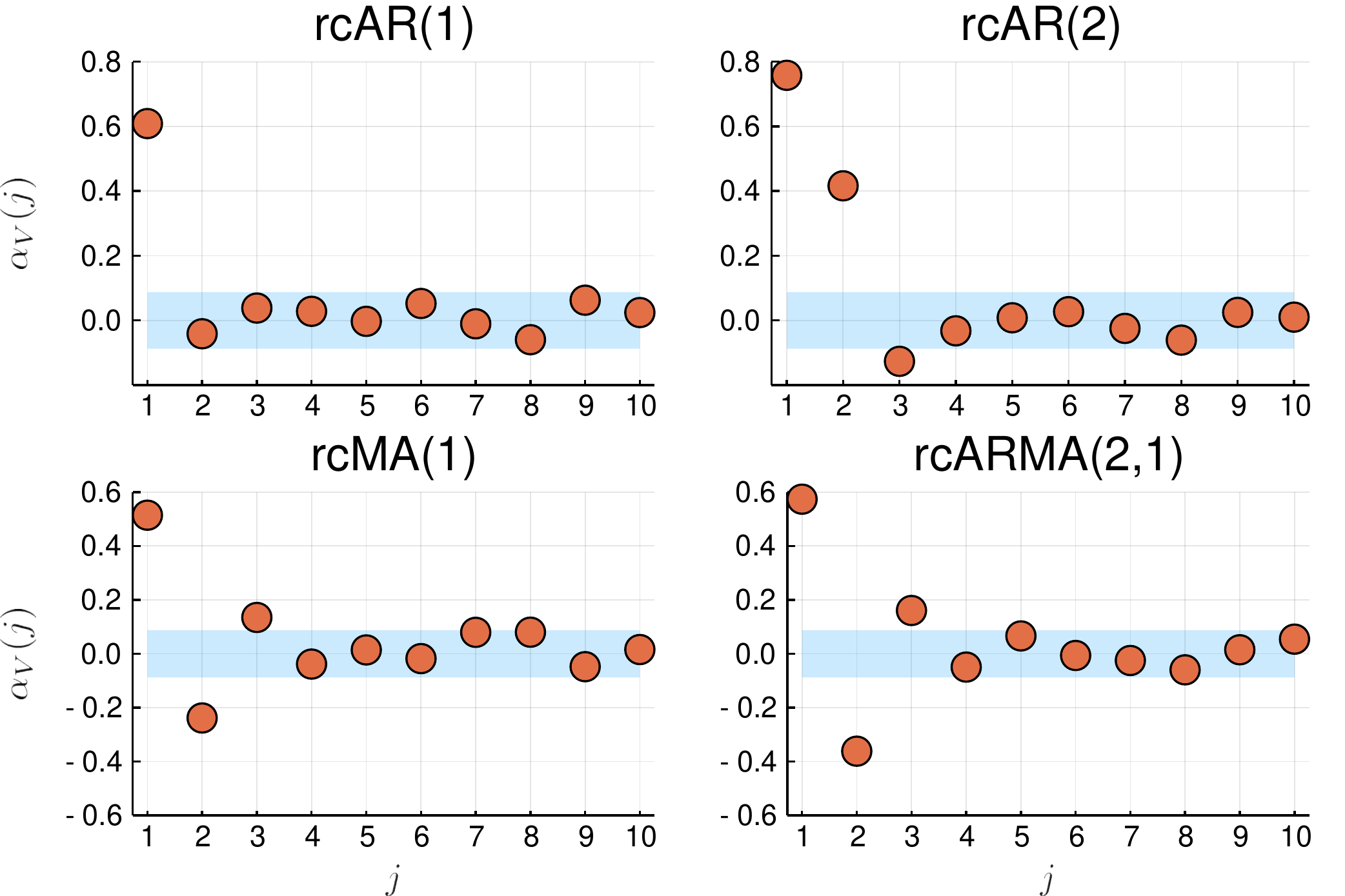}
\caption{Partial autocorrelation estimated from i.i.d. rcAR(1) with
$\Phi_k\deq \mathcal U(0,1)$, rcAR(2) with $\Phi^1_k\deq \mathcal U(0,1),
\Phi^2_k\deq \mathcal U(0,3/4)$, rcMA(1) with $\Theta_k\deq \mathcal U(0,2)$,
and rcARMA(2,1) with $\Phi^1_k\deq \mathcal U(0,1)$, $\Phi^2_k\deq \mathcal
U(-1/2,0)$, $\Theta_k\deq \mathcal U(0,2)$. Blue areas are 95\% confidence
intervals around 0. The memory function nicely fits the predictions, in particular
one can easily distinguish the simplest rcAR(1) model. The choice of the presented
rcARMA(2,1) was motivated by the Langevin equation \Ref{eq:hp}. Here we used a
500 point trajectory---to observe only the characteristic rcAR(1) behaviour
even shorter, 200-300 point trajectories are sufficient.}
\label{fig:pacf}
\end{figure}

For the i.i.d. rcAR(1) the statistical procedure becomes very simple: just
estimate $\alpha_X$ and check if it has a significant non-zero value only for
$j=1$. This value is simply $\alpha_X(1)=\mathbb{E}\left[\Phi_k\right]$. For
an illustration see figure \ref{fig:pacf}. For higher order AR processes
there are also exact equations linking the values of partial autocorrelation,
covariance and coefficients of the model \cite{BJ,BD}.

\begin{figure}
\centering
\includegraphics[width=7.6cm]{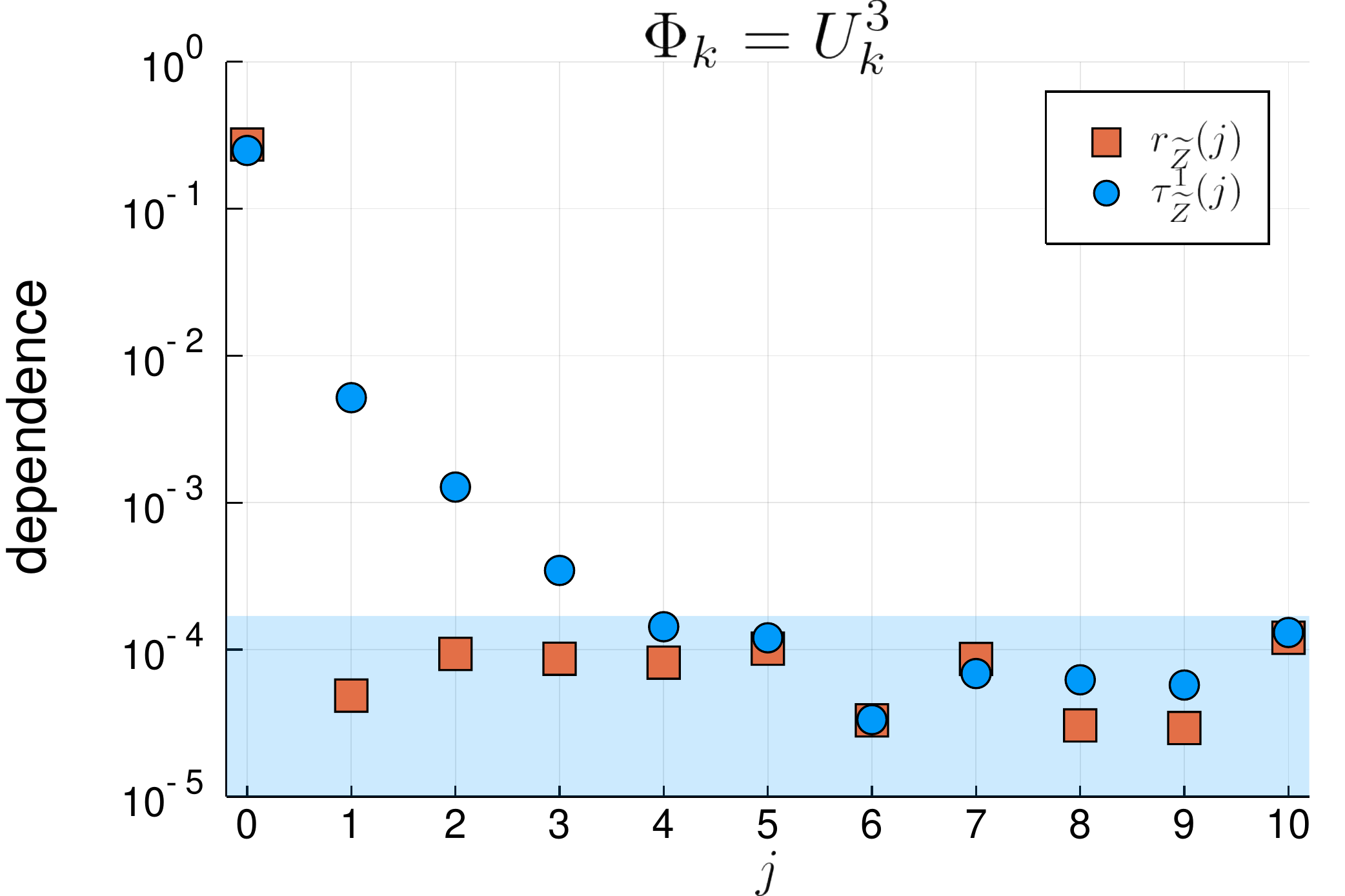}
\includegraphics[width=7.6cm]{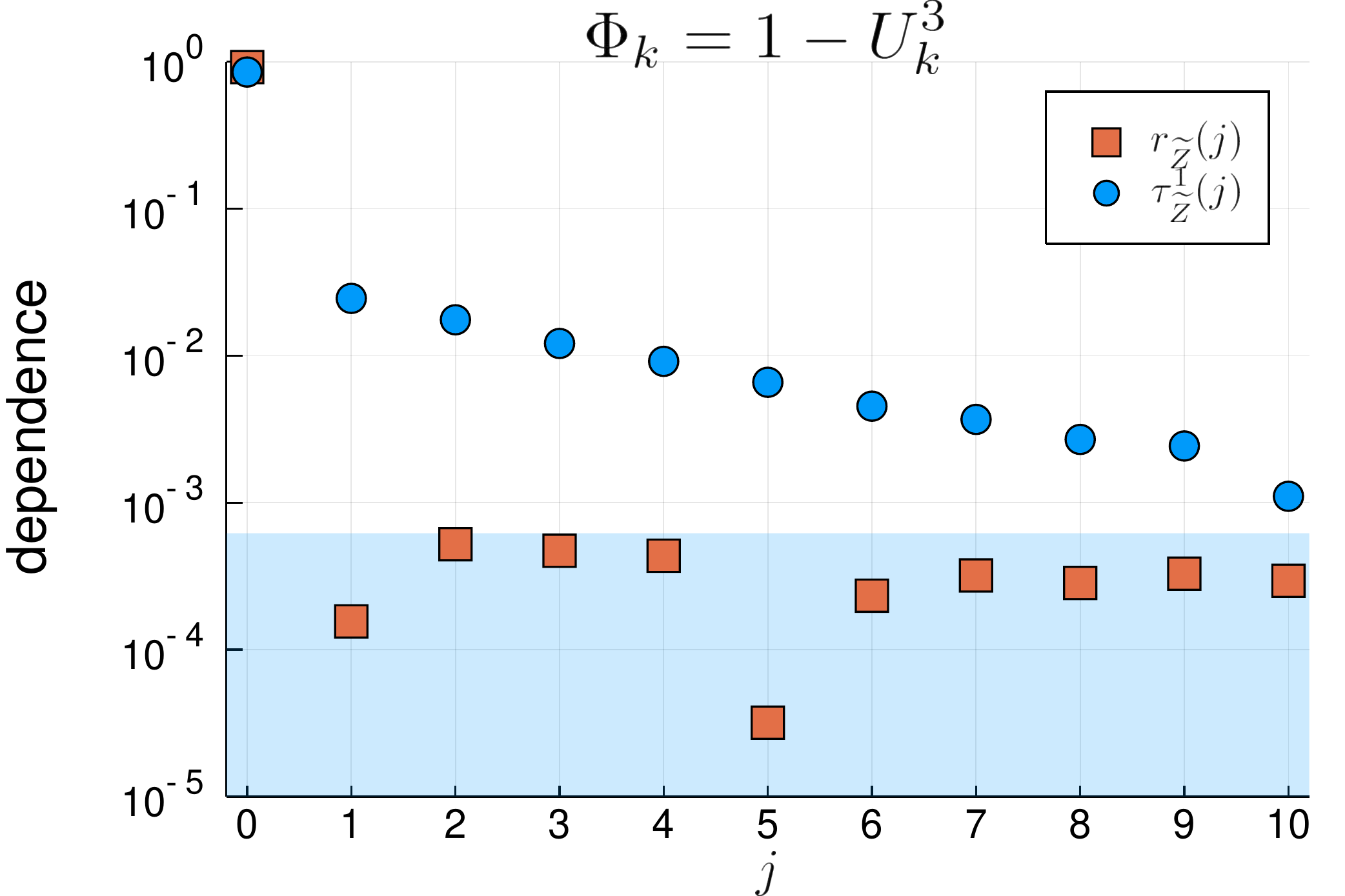}
\caption{Comparison of the covariance  and codifference for the series
$\widetilde Z_k=V_k-\mathbb{E}\left[\Phi_k\right] V_{k-1}$ generated with
two different i.i.d. $\Phi_k$ obtained from $U_k\deq\mathcal U(0,1)$ and
$\Theta_k=\sqrt{\Phi_k}$. Blue areas are 95\% confidence intervals around
0. The covariance is statistically zero for $j>0$, the codifference exhibits an
exponential decay. We used two extremely long $10^7$-point trajectories in
order to present a long range of $j$ and very weak dependence (statistically
significant even around $10^{-3}$). Much shorter trajectories are sufficient
for less extreme cases, the requirements in general are similar to those
for the covariance estimation.}
\label{fig:covCdfComp}
\end{figure}

We now proceed to show that there actually is a way to detect differences between
the dependence structure of ARMA and rcARMA. First, we estimate the mean rcAR(1)
coefficient $\varphi=\mathbb{E}\left[\Phi_k\right]$, for instance,  by using the
partial autocorrelation. Next, the typical method of statistical verification for
AR processes is to consider 
\begin{equation}
\label{eq:ZtildeDef}
\widetilde Z_k\defeq V_k-\varphi V_{k-1}.
\end{equation}
For the AR(1) with deterministic coefficient $\phi=\varphi$ this would be an estimate of the noise $Z_k$. Therefore, one could then
use many of the testing methods if the series is i.i.d. However, for the rcAR(1),
\begin{equation}
\widetilde Z_{k}=(\Phi_k-\varphi)V_{k-1}+Z_k.
\end{equation}
As $\mathbb{E}\left[\Phi_k-\varphi\right]=0$ the series is still uncorrelated,
$\mathbb{E}[\widetilde Z_k \widetilde Z_{k+1}]=0$. It is a natural
consequence of the applied procedure, which fits the system by removing the
linear dependence. But the series $\widetilde Z_k$ remains non-linearly
dependent. The value $\widetilde Z_k$ still depends on $V_{k-1}$ and consequently
also on $\Phi_{k-1}$ appearing in the expression for $\widetilde Z_{k-1}$. This
purely non-linear dependence can be detected only using non-linear measures of
memory. The codifference function (see \cite{jakub} for a detailed discussion) is
one possible choice. It was proposed as a tool to study $\alpha$-stable
variables, because it is finite even for variables with no second, or even
lower, moments \cite{taqqu}---however its usefulness is by no means limited
to this class. One of the few possible close definitions is the formula
\begin{equation}
\label{eq:cdfDef}
\tau_X^\theta(j)\defeq\f{1}{\theta^2}\ln\f{\mathbb{E}\left[\e^{\mathrm{i}\theta(X_{j}
-X_0)}\right]}{\mathbb{E}\left[\e^{\mathrm{i}\theta X_{j}}\right]\mathbb{E}\left[\e^{
-\mathrm{i}\theta X_0}\right]}.
\end{equation}
For any Gaussian process the codifference equals the covariance, so the difference
between the two indicate any non-Gaussianity of the PDF and, what is important
here, of the memory structure. For i.i.d. rcAR it can be expressed as a
function of coefficients, and it can be analytically approximated, see
\Ref{eq:cod1iid} and the derivation below.
As a simple example let us take the uniformly distributed $\Phi_k\deq\mathcal
U(0,1)$ and
$\Theta_k=\sqrt{D_k\Phi_k},D_k\deq\mathcal U(0,1)$. The covariance
is zero as $\mathbb{E}[\widetilde Z_k\widetilde Z_{+1}]=0$,
but the simulation yields $\tau_{\widetilde Z}^1(1)\approx 0.0023$, in
good agreement with the analytic approximation is close $412/170535=0.0024$.
Two more examples of rcAR(1) with higher dispersion of i.i.d. coefficients and
even more prominent non-linear memory are shown in figure \ref{fig:covCdfComp}.

\subsection{Non-Gaussianity}

One possible approach of visualising the non-Gaussian behaviour of a diffusive
process is to quantify the non-Gaussian nature of its dispersion in a manner
similar to how we quantified non-Gaussian memory in the last section. We propose to
use the non-linear measure of dispersion \cite{jakub}
\begin{equation}
\label{eq:LCF}
\zeta^\theta_X(j)\defeq-\f{2}{\theta^2}\ln\left(\mathbb{E}\left[\cos(\theta X_j)
\right]\right),
\end{equation}
which is called log-characteristic function (LCF). It has the property that if
the process is Gaussian it equals the MSD for any $\theta$. Therefore estimating
it and plotting it together with the MSD reveals the non-Gaussian nature of the
observed process. As the MSD averages over squares of the data, small values
become even smaller and larger values become even more emphasised. Thus the
MSD gives a large importance to the spread of extremal values. Conversely,
for the LCF large values are translated into rapid oscillations of the cosine
function and cancel each other, the LCF measures mainly the spread of the bulk.

For the classical case of non-Gaussian normal diffusion \Ref{eq:LaplPDF} the
LCF is linear for small times, but logarithmic for large times, so it gets
dominated by the linear MSD. This insight is more general. Random coefficient
models discussed in this work can only describe diffusion processes with tails
thicker than Gaussian. This means that compared to Gaussian diffusion with the
same MSD more mass is located in the tails than in the bulk, and the LCF must
be smaller than the MSD.

\begin{figure}
\centering
\includegraphics[width=7.0cm]{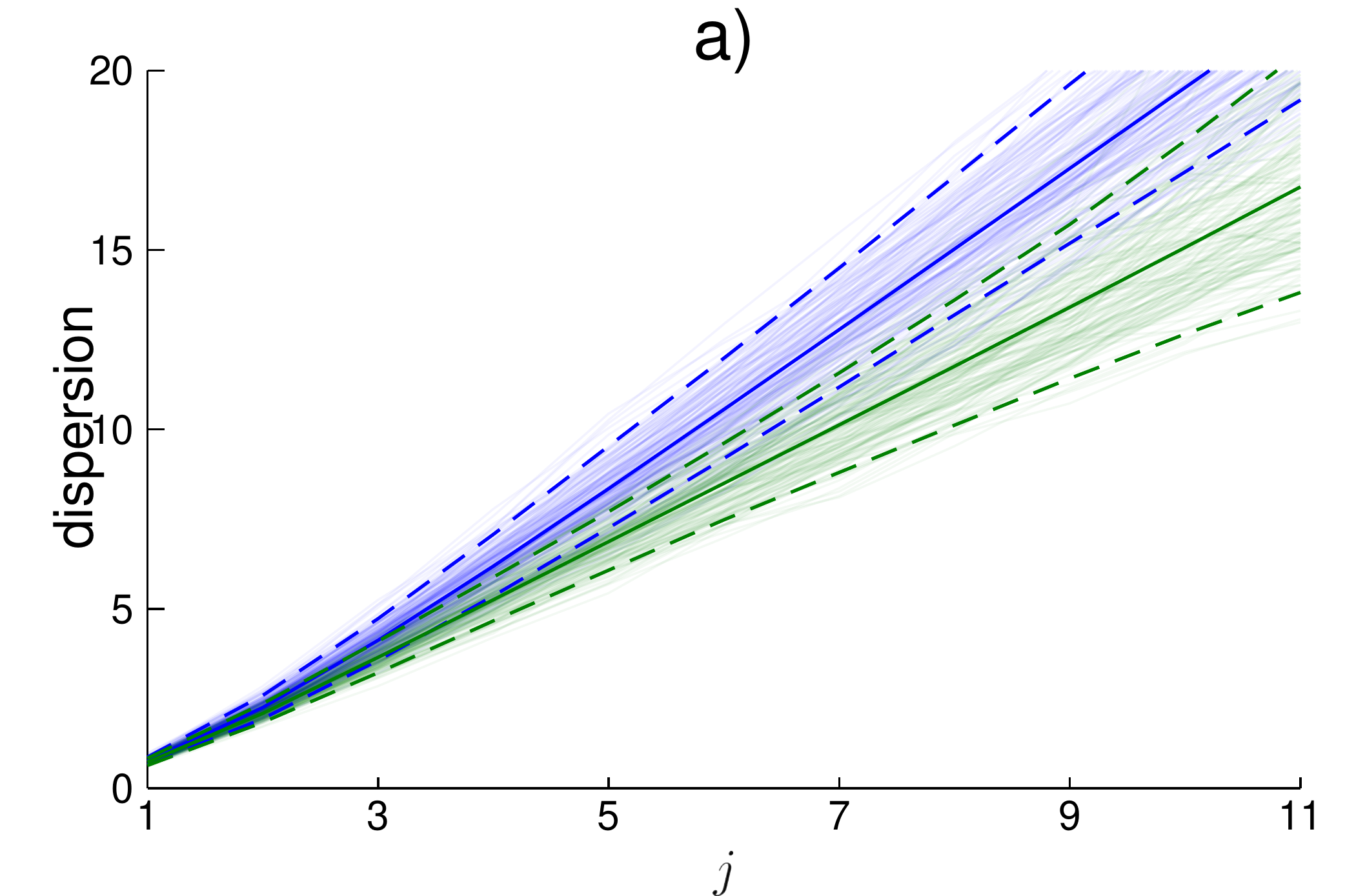}
\includegraphics[width=7.0cm]{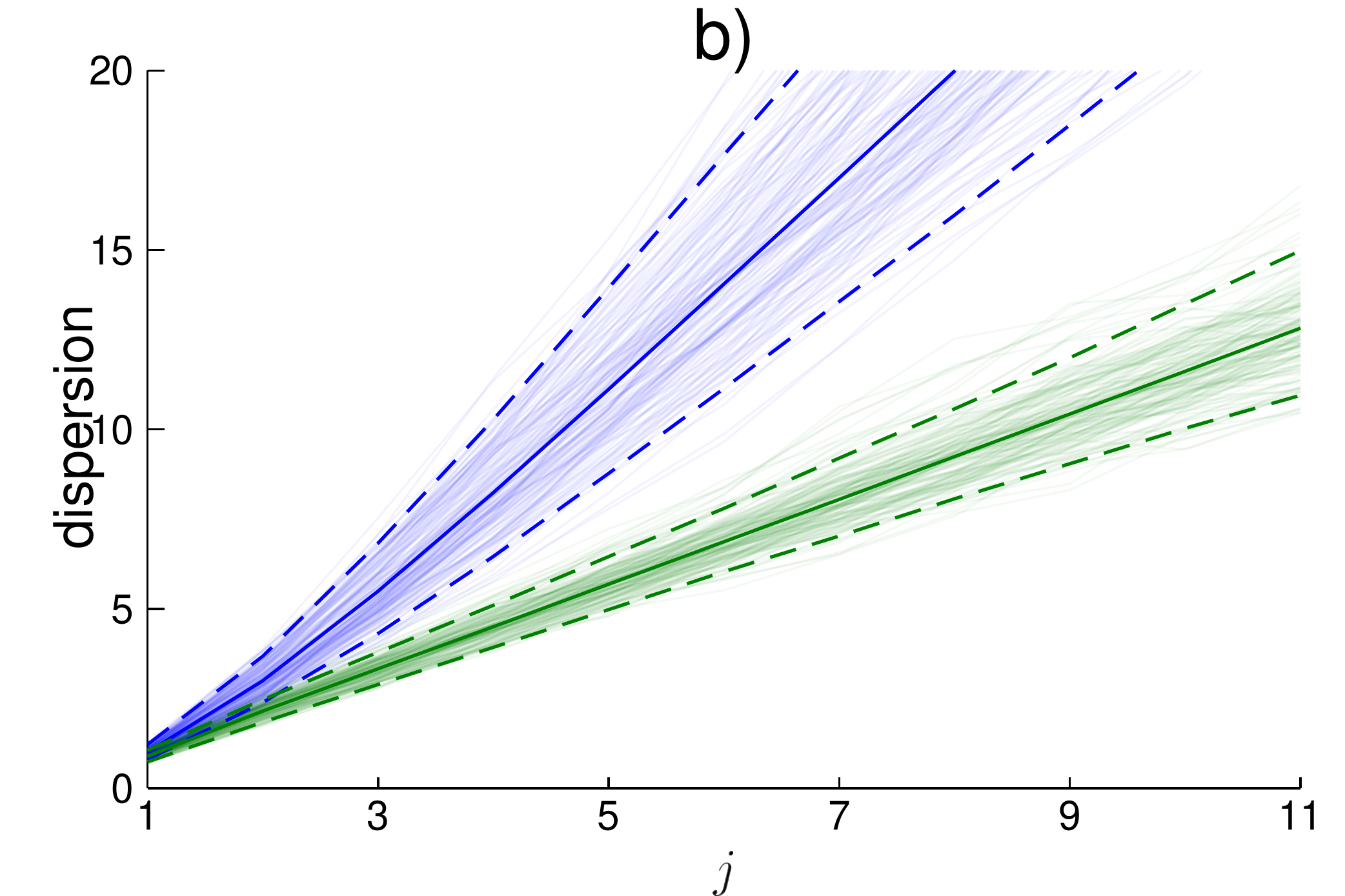}\\
\includegraphics[width=7.0cm]{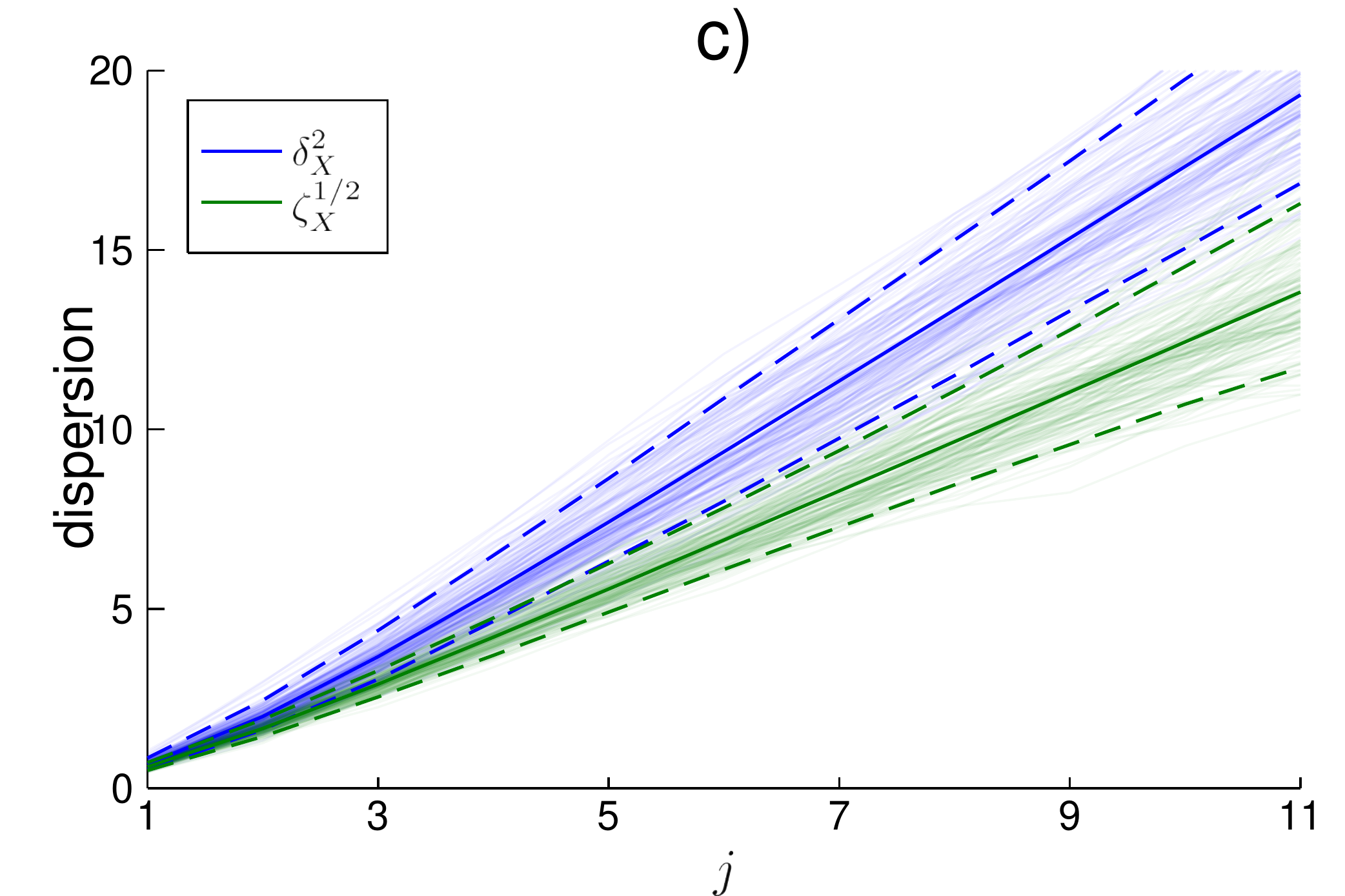}
\caption{LCF and MSD calculated from 500 samples of the integrated
rcAR processes. Semi-transparent lines are the individual estimated
LCFs and MSDs, solid lines are the averaged ones, dashed lines
denote $90\%$ confidence intervals. Case a) is $\Phi_k\deq\mathcal
U(0,1), \Theta_k=\sqrt{\Phi_k}$; case b) is $\mathrm{P}(\Phi_k =
0) = 1/2 = \mathrm{P}(\Phi_k=1),\Theta_k=\sqrt{\Phi_k}$; case c) is
$\Phi_k=1/2,\Theta_k\deq 1/2\times\mathcal E(1)$, where $\mathcal E$ is
an exponential distribution. Cases a) and b) are based
on variants of i.i.d. rcAR considered before, case c) is an example of a
motion with varying diffusivity strongly concentrated at 0.}
\label{fig:LCF}
\end{figure}

To use the LCF in practice one needs to fix the parameter $\theta$. For too
small values the LCF converges to the MSD because $\cos(\theta x)\approx 1-\frac{
1}{2}\theta^2x^2$, for too large values the estimation becomes challenging, because
$\ln(0)$ diverges. In a typical case reasonable values are those for which $\theta
\mathbb{E}\left[|X(t)|\right]\approx\pi$ so that $X(t)$ explores mainly the first
period of the cosine function. In figure \ref{fig:LCF} we show a practical
application of this technique performed using a relatively small sample of 500
integrated rcAR trajectories.

For rigorous statistical testing of the Gaussianity we propose to use the
standard Jarque-Bera (JB) test \cite{jbtest87}. It is a goodness of fit
test of whether sample data have skewness and kurtosis that match a normal
distribution. For the sample $\{x_1, \ldots, x_n\}$ of observation the JB
statistic is defined as
\begin{equation}
J=\frac{n}{6}\left(\bar{s}^2+\frac{\bar{\kappa}^2}{4}\right),
\label{eqn:jb_stat}
\end{equation}
where $\bar{s}$ is the sample skewness and $\bar{\kappa}$ is the sample excess kurtosis.
Samples from a normal distribution have an expected skewness of $0$ and an
expected kurtosis of $3$ (excess kurtosis 0). Any deviation
from this increases the JB statistic. The test is considered as standard and is
implemented in various numerical packages, such as R \texttt{tseries}, Python
\texttt{SciPy.stats}, Julia \texttt{HypothesisTests}, or Matlab.

This test is considered as one of the most powerful tests on normality but, if
non-Gaussianity is detected, it does provide information on the origins of this
behaviour. However, the idea of observing the kurtosis leads to an algorithm to
distinguish between Gaussian and non-Gaussian (in particular, L{\'e}vy stable)
distributions \cite{buretalkurt2012}. It is based on the empirical cumulative
excess kurtosis (ECEK) which is defined as follows,
\begin{equation}
\label{eqn-ecfm}
K(k)=\frac{\frac{1}{k}\sum_{i=1}^k(x_i-\bar{x})^4}{\left(\frac{1}{k}\sum_{i=1}^k
(x_i-\bar{x})^2\right)^2}-3,
\end{equation}
where $\bar{x}$ is an arithmetic mean of the random sample. This simple statistic
serves as an indicator for whether there is a noticeable difference between Gaussian
and non-Gaussian distributions. For the Gaussian case, for large numbers of
observations it converges to the theoretical excess kurtosis 0, while for the
non-Gaussian case the ECEK does not tend to 0 with increasing number $k$ of
observations and, moreover, for distributions that do not have a finite forth
moment it does not converge at all and behaves chaotically \cite{buretalkurt2012}.

To distinguish between Gaussian and non-Gaussian distributions we also advocate
the application of the discrimination algorithm based on examining the rate
of convergence to the Gaussian law by means of the central limit theorem
\cite{buretalclt2015}. The idea of the algorithm is to analyse the convergence
of the estimated index $\alpha$ of stability for sequential bootstrapped
samples from the analysed data. If the estimated values converge to 2,
then the data are light-tailed and belong to the domain of attraction of
the Gaussian law. In particular, if the data are Gaussian, the estimated
values should be equal to 2 for most of the cases. If the data belong to
the domain of attraction of a non-Gaussian L{\'e}vy stable law, then the values
should converge to a constant less than 2.

\begin{figure}
\centering
\includegraphics[width=16cm]{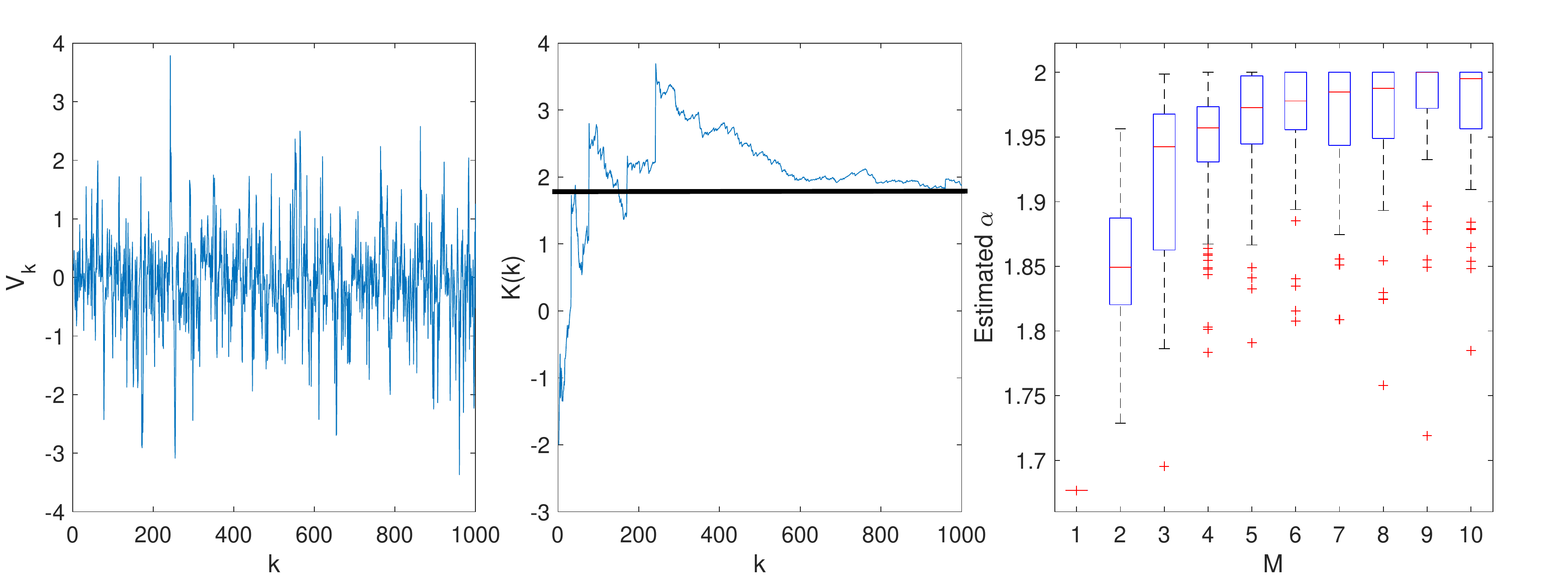}
\caption{Left: Trajectory of $V_k$ for which $\Phi_k\deq\mathcal U(0,0.95)$ and
$\Theta_k=\sqrt{\Phi_k}$. Middle: Empirical cumulative excess kurtosis for the
simulated trajectory. The solid black line represents the true kurtosis value
which is equal to $1.83$. Right: Estimated values of the index of stability
$\alpha$. The boxplots (for block lengths $M \in [1,10]$) constructed from 100
bootstrapped samples each of length 1000. The obtained results clearly suggest
that the simulated trajectory exhibits a non-Gaussian behaviour and its
distribution is leptokurtic but belongs to the normal domain of attraction.}
\label{fig:boxplots}
\end{figure}

To illustrate the usefulness of the above statistical tools we consider an rcAR
process for which the $\Phi_k$ are independent uniformly distributed on the
interval $(0,0.95)$ and $\Theta_k=\sqrt{\Phi_k}$. In the left panel of
figure \ref{fig:boxplots} we can see a simulated trajectory of $V_k$ of
length 1000 points. In the middle panel we display a plot of the ECEK
for the simulated trajectory. We can see that the function converges to a
constant close to 2 which clearly indicates a non-Gaussian behaviour and a
finite positive excess kurtosis of the underlying distribution (a leptokurtic
distribution). We also calculated analytically the exact value of the excess
kurtosis (equations \Ref{eq:kurt} and \Ref{eq:S4}) and obtained the value $1.83$,
which coincides with the final values of the ECEK function. Finally, in
the right panel of figure \ref{fig:boxplots} we show the estimated values
of the stability index $\alpha$, for different non-overlapping consecutive
blocks of length $M$. For the first sample ($M=1$; corresponding to the
whole trajectory), the value is significantly lower than 2, but for the other
$M$ the values increase and tend to 2. This suggests that the simulated
trajectory exhibits a non-Gaussian behaviour but its distribution belongs
to the normal domain of attraction. The obtained results clearly suggest
that the simulated trajectory exhibits a non-Gaussian behaviour and the
underlying distribution is heavier-tailed than Gaussian (is leptokurtic)
but belongs to the normal domain of attraction. The non-Gaussianity is also
confirmed by the JB test which returns a $p$-value less than $0.001$.

\subsection{Parameter estimation for rcAR processes}

Finding reliable estimators of the parameters of the rcARMA models is a very
important challenge. In this section we present preliminary ideas for some
special cases of the models.

For an rcAR process with $\Phi$ being a simple function one can estimate
its parameters either by extracting the constant periods of $\Phi$ and
applying classical fitting techniques for the autoregressive processes to
the extracted parts, or to consider a method which directly incorporates the
information on switching between different autoregressive processes.

\begin{figure}
\begin{center}
\includegraphics[width=8cm]{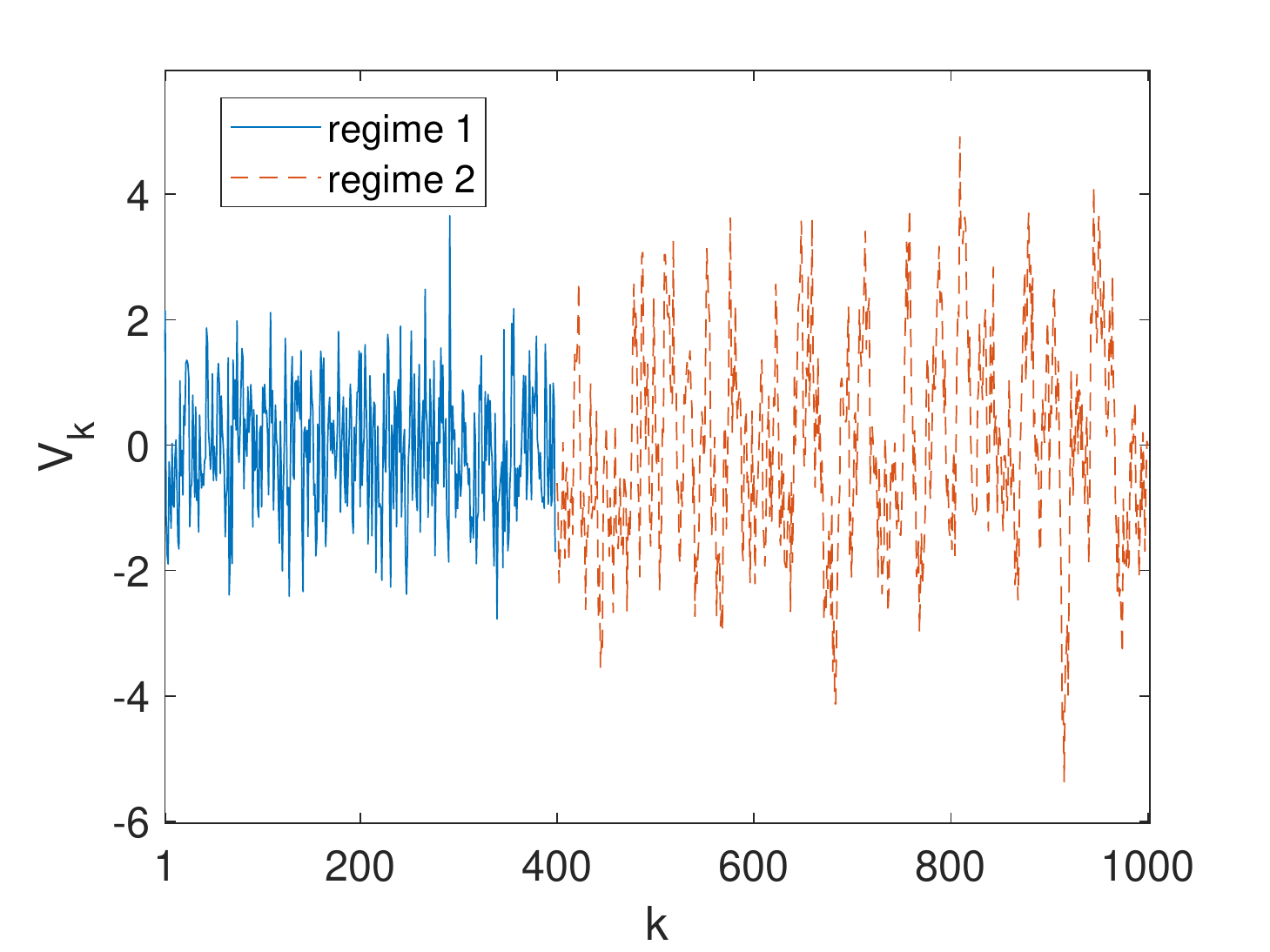}
\caption{Two regimes identified by Hidden Markov Model methodology for a
trajectory of $V_k$ for which $\Phi_\alpha$ changes from 0.2 to 0.8 at point
400. The estimated regime switching point is $399$ which is very close to
the true value.}
\label{fig:HMM}
\end{center}
\end{figure}

We now follow the latter idea and apply an algorithm based on hidden Markov
Models (HMMs) \cite{Stan_19,baum1966,sungetal2017}. This algorithm assumes
that the trajectories switch among discrete diffusive states according to
a stochastic (Markov) process. We consider a modification of the classical
HMM, where the Markov regime switching is combined with  AR$(1)$ processes
\cite{Janczura2012}. To show the usefulness of this technique we take
into account a two-regime parameter switching model, that is, a model with
both regimes driven by AR$(1)$ processes with the autoregressive parameters
$\phi_1$ and $\phi_2$. In figure \ref{fig:HMM} we present a simulated
trajectory of $V_k$ of length 1000 for which $\Phi_\alpha$ changes
from 0.2 to 0.8 at point 400. The estimated regime switching point is
$399$ which is very close to the true value, and the estimated $\phi_1$ and
$\phi_2$ coefficients are $0.17$, $0.79$ which confirms the usefulness of the
procedure.

\section{Discussion}

As modern experiments such as single particle tracking routinely produce large
amounts of data for the thermally and actively driven motion of test particles,
the extraction of physical information from the garnered data becomes ever more
important. On the one hand, this is met by the analysis of a growing number of
complementary observables such as time averaged MSD \cite{he,johannes}, higher
order moments or mean maximal statistics \cite{vincent1}, p-variation methods
\cite{marcin1}, first-passage methods \cite{olivierpnas}, or single
trajectory power spectra \cite{krapf,gleb}. On the other hand objective methods
such as maximum likelihood approaches \cite{remi,michael,samu,samu1} or machine
learning \cite{gorka} are being recognised as useful tools. Here the goal is to
identify the physical model giving rise to the observed motion and estimate the
involved parameters.

An alternative method, well established in financial market studies, is time
series analysis. The latter seemingly does not share a lot with the physical
approach of modelling. It puts the emphasis not on explaining the observed systems,
but instead concentrates on fitting and prediction. Given a class of discrete
linear filters suited to describe the general features of the data (for us
the non-Gaussianity and linear MSD) it offers a wide choice of powerful methods
for finding the optimal model and validating it. In other words by design it
concentrates on the phenomenological description.

While a pure time series analysis approach initially may be discouraging for
physically-minded researchers, the undeniable effectiveness of the time series
methods make any connections to the framework of physics interesting and
important. As noted in section \ref{s:physDer} the correspondence between the
Gaussian Langevin equation and the ARMA process is known (in fact, it was
established as early as in 1959 \cite{phillips}), but it never seemed to have
gained a widespread appeal in the physical or biological physics community.
Nevertheless, some recent progress was made, especially in topics related to
the identification of fractional dynamics \cite{Balcerek_19,burnecki19} and
accounting for the distortions made by measurement devices
\cite{drobczynskiSlezak, sikora17}.

This paper continues this line of research and aims at promoting to sample the
best of both worlds. This is hard to achieve on a general level, but we study
here important cases when a fruitful compromise can be made, and new information
provided in the challenging analysis of non-Gaussian diffusion. Omitting the
dynamics at time scales significantly shorter than what is available from
observation (shorter than the sampling time $\Delta t$) leads to conceptually
simple autoregressive models. These are straightforward and quick to simulate
(using a medium class computer we were generating ten million values in under
2 seconds) and they provide a wide choice of statistical methods. Some of those,
such as the partial autocorrelation, are intrinsically linked to discrete
dynamics and have no continuous-time counterpart. Others, such as the kurtosis
and codifference, profit from the analytic methods available for discrete variables.

In section \ref{sec:stat} we illustrated important properties of the rcARMA
processes and provided a list of statistical tools that can be useful in their
identification and validation. In particular we showed that the
codifference function can be used to distinguish between ARMA and rcARMA
processes. To illustrate the non-Gaussianity of the rcARMA models we considered
a non-linear measure of dispersion, namely the log-characteristic function and
the empirical cumulative excess kurtosis, and studied the domain of attraction
of the underlying distribution. Finally, we presented a hidden Markov model
methodology which can be useful in fitting the process parameters. We also
stress that the presented tools can be successfully applied for quite moderate
lengths of the trajectories.

Our considerations are general and not limited to any particular system, as we
discuss for related models based on random coefficient autoregression, which
stem from a widely used form of the Langevin equation and reasonable models
of heterogeneous environment. Thus, the provided methods are also universal.
We hope that this study will promote the use of autoregressive models in
modern data analysis, as well as prompt further studies into the physical
meaning of these models.

\ack

J\'S acknowledges support through the Polish National Science Centre grant
2016/22/M/ST1/00233. KB acknowledges support through Beethoven Grant No.
DFG-NCN 2016/23/G/ST1/04083. RM acknowledges funding from DFG, grants ME1535/6-1
and ME1535/7-1. RM was supported by an Alexander von Humboldt Polish Honorary
Research Scholarship from the Foundation for Polish Science (Fundacja na rzecz
Nauki Polski).

\appendix

\section{Appendix: Moments of $S$ and the approximation of codifference}

Here we show the calculation of $\mathbb{E}\left[S^4\right]$ and the approximation
of the codifference for an rcAR with i.i.d. coefficients. Starting with \Ref{eq:S2}
we write
\begin{equation}
\mathbb{E}\left[S^4\right]=\mathbb{E}\left[\left(\Theta_k^2+\Phi_k^2\Theta_{k-1}^2
+\Phi_k^2\Phi_{k-1}^2\Theta_{k-2}^2+\Phi_k^2\Phi_{k-1}^2\Phi_{k-1}^2\Theta_{k-3}^2
+\ldots\right)^2\right].
\end{equation}
We express this square as the sum of the elements squared plus twice the sum of
each element multiplied by all elements to the right of it. The sum of squares
is simply $\mathbb{E}\left[\Theta_k^4\right]/(1-\mathbb{E}\left[\Phi_k^4\right])$.
As for the rest, in each element a factor of type $\mathbb{E}\left[\Phi_{k-i}^2
\Theta_{k-i}^2\right]\mathbb{E}\left[\Theta_{k-j}^2\right]= \mathbb{E}\left[\Phi_
k^2\Theta_k^2\right]\mathbb{E}\left[\Theta_k^2\right]$ appears. Note that it does
not necessarily decouple, because in general $\Phi_k$ and $\Theta_k$ for fixed $k$
may be dependent. The remaining factor is 
\begin{align}
&1&+&\mathbb{E}\left[\Phi_k^2\right]&+&(\mathbb{E}\left[\Phi_k^2\right])^2&+&(
\mathbb{E}\left[\Phi_k^2\right])^3&+&\ldots \nonumber\\
&\mathbb{E}\left[\Phi_k^4\right]&+&\mathbb{E}\left[\Phi_k^4\right]\mathbb{E}
\left[\Phi_k^2\right]&+&\mathbb{E}\left[\Phi_k^4\right](\mathbb{E}\left[\Phi_k^2
\right])^2&+&\mathbb{E}\left[\Phi_k^4\right](\mathbb{E}\left[\Phi_k^2\right])^3&
+&\ldots\nonumber\\
&(\mathbb{E}\left[\Phi_k^4\right])^2&+&(\mathbb{E}\left[\Phi_k^4\right])^2
\mathbb{E}\left[\Phi_k^2\right]&+&(\mathbb{E}\left[\Phi_k^4\right])^2(\mathbb{E}
\left[\Phi_k^2\right])^2&+&(\mathbb{E}\left[\Phi_k^4\right])^2(\mathbb{E}\left[
\Phi_k^2\right])^3&+&\ldots\nonumber\\
&(\mathbb{E}\left[\Phi_k^4\right])^3&+&(\mathbb{E}\left[\Phi_k^4\right])^3
\mathbb{E}\left[\Phi_k^2\right]&+&(\mathbb{E}\left[\Phi_k^4\right])^3(\mathbb{
E}\left[\Phi_k^2\right])^2&+&(\mathbb{E}\left[\Phi_k^4\right])^3(\mathbb{E}\left[
\Phi_k^2\right])^3&+&\ldots\nonumber\\
&\quad \vdots 
\end{align}
We recognise the product of two geometric series in the above. Finally, the fourth
moment becomes
\begin{equation}
\label{eq:S4}
\mathbb{E}\left[S^4\right]=\f{\mathbb{E}\left[\Theta_k^4\right]}{1-\mathbb{E}
\left[\Phi_k^4\right]}+ 2\mathbb{E}\left[\Phi_{k}^2\Theta_{k}^2\right]\mathbb{E}
\left[\Theta_k^2\right]\f{1}{1-\mathbb{E}\left[\Phi_k^4\right]}\f{1}{1-\mathbb{E}
\left[\Phi_k^2\right]}.
\end{equation}

Given these moments, one can easily calculate the kurtosis. It also makes it
possible to obtain an analytic approximation of the codifference of the estimated
noise $\widetilde Z_k\defeq V_k-\varphi V_{k-1},\varphi=\mathbb{E}\left[\Phi_k
\right]$. We show the procedure for $\tau_{\widetilde Z}^\theta(1)$. The method
for other $\tau_{\widetilde Z}^\theta(j),j >1$ would be the same. We denote
$\widetilde\Phi_k\defeq\Phi_k-\varphi$ and start with a straightforward
conditioning which yields
\begin{equation}
\label{eq:cod1iid}
\tau_{\widetilde Z}^\theta(1)=\f{1}{\theta^2}\ln\f{\mathbb{E}\left[\exp\left(
-\f{\theta^2}{2}\left((\widetilde\Phi_k\Phi_{k-1}-\widetilde\Phi_{k-1})^2S^2
+(\widetilde\Phi_k-1)^2\Theta_{k-1}^2+\Theta_k^2 \right)\right)\right]}{
\mathbb{E}\left[\exp\left(-\f{\theta^2}{2}\left(\widetilde\Phi_k^2S^2+\Theta_k
^2\right)\right)\right]^2},
\end{equation}
where $S^2$ is assumed to have the distribution as in \Ref{eq:S2} and be
independent from $\Phi_k,\Theta_k,\Phi_{k-1},\Theta_{k-1}$. Expanding the
exponents in a Taylor series yields an approximation for the codifference
which depends only on the moments of $\Phi_k$ and $\Theta_k$. Expansion up
to first order, $\e^{-x}\approx 1-x$, is equivalent to calculating the covariance,
so it is zero. The second order, $\e^{-x}\approx1-x+x^2/2$ in general already
provides quite good non-zero estimate. For the numerator it is
\begin{align}
&\approx1-\f{\theta^2}{2}\left(\mathbb{E}\left[\left(\widetilde\Phi_k\Phi_{k-1}
-\widetilde\Phi_{k-1}\right)^2\right]\mathbb{E}\left[S^2\right]+\mathbb{E}\left[
(\widetilde\Phi_k-1)^2\Theta^2_{k-1}+\Theta_k^2\right]\right)\nonumber\\
& +\f{\theta^4}{8}\left(\mathbb{E}\left[\left(\widetilde\Phi_k\Phi_{k-1}
-\widetilde\Phi_{k-1}\right)^4\right]\mathbb{E}\left[S^4\right]\right.\nonumber\\
&+2\mathbb{E}
\left[\left(\widetilde\Phi_k\Phi_{k-1}-\widetilde\Phi_{k-1}\right)^2\left(
(\widetilde\Phi_k-1)^2\Theta^2_{k-1}+\Theta_k^2\right)\right]\mathbb{E}\left[
S^2\right]\nonumber\\
&+\left.\mathbb{E}\left[\left((\widetilde\Phi_k-1)^2\Theta^2_{k-1}+\Theta_k^2
\right)^2\right]\right).
\end{align}
The corresponding denominator is expanded into
\begin{align}
&\approx 1-{\theta^2}\left(\mathbb{E}\big[\widetilde\Phi_k^2\big]\mathbb{E}
\left[S^2\right]+\mathbb{E}[\Theta_k^2]\right)+\f{\theta^4}{4}\Bigg(
\left(\mathbb{E}\big[\widetilde\Phi_k^2\big]\mathbb{E}\left[S^2\right]
+\mathbb{E}\left[\Theta_k^2\right]\right)^2\nonumber\\
&+\left(\mathbb{E}\big[\widetilde\Phi_k^4\big]\mathbb{E}\left[S^4
\right]+2\mathbb{E}\big[\widetilde\Phi_k^2\Theta_k^2\big]\mathbb{E}\left[
S^2\right]+\mathbb{E}\left[\Theta^4_k\right]\right)\Bigg).
\end{align}
The expected values that appear in this formulas are just linear combinations
of the first four moments of $\Phi_k, \Theta_k$ and their products. For
variables generated as powers of a uniform distribution these are just
integrals over polynomials. Also, without significant change the logarithm
can be replaced by the fraction $\ln(x)\approx x-1$, which leads to a
formula being a fraction of coefficient moments up to the fourth order.

\end{document}